\documentclass[preprint,12pt]{elsarticle}

\journal{Astronomy and Computing}

\newcommand{\ghlink}{\href{https://github.com/emilleishida/CorrMatrix_ABC}{\faGithub}}
\newcommand{\nblink}[1]{\href{https://github.com/emilleishida/CorrMatrix_ABC/blob/master/scripts/#1.ipynb}{\faFileCodeO}}

\usepackage{amsmath}
\usepackage{graphicx}	% Including figure files
\usepackage{enumerate}
\usepackage{color, colortbl}
\usepackage{natbib}
\usepackage[]{algorithm2e}
\usepackage{amssymb}	% Extra maths symbols
\usepackage{bm}
\usepackage{ulem}
\usepackage{multirow}
\usepackage{comment}
\usepackage{hyperref}
\usepackage{fontawesome}
\usepackage[utf8]{inputenc}

\usepackage{sidecap}
\sidecaptionvpos{figure}{c} %put side captions in center
\usepackage{float}

\newcommand{\tr}{{\rm tr\,}}

\newcommand{\Z}{_{\text{Z}}}
\newcommand{\SIM}{\text{sim}}
\newcommand{\OBS}{\text{obs}}
\newcommand{\Sim}{_{\SIM}}
\newcommand{\Obs}{_{\OBS}}

\definecolor{orange}{RGB}{255,127,0}
\definecolor{grey}{RGB}{200,200,200}
\definecolor{darkgreen}{rgb}{0.1,0.5,0.1}

\newcommand{\mat}[1]{\bm{\mathrm{#1}}}
\renewcommand{\vec}[1]{\bm{#1}}
\newcommand{\LG}{\log_{10}}

\newcommand{\hi}{\cellcolor{grey}}

\usepackage[margin=2.2cm]{geometry}

\begin{document}

\begin{frontmatter}

\title{Sidestepping the inversion of the weak-lensing covariance matrix with Approximate Bayesian Computation}

\author[A]{Martin Kilbinger}
\affiliation[A]{
    addressline={AIM, CEA, CNRS, Universit\'e Paris-Saclay,
    Universit\'e de Paris},
    city={Gif-sur-Yvette},
    postcode={91191},
    country={France}
}

\author[B]{Emille E.~O. Ishida}
\affiliation[B]{
    addressline={Universit\'e Clermont Auvergne, CNRS/IN2P3, LPC},
    postcode={63000},
    city={Clermont-Ferrand},
    country={France}
}

\author[C]{Jessi Cisewski-Kehe}
\affiliation[C]{
    addressline={Department of Statistics, University of Wisconsin},
    city={Madison},
    postcode={53706},
    country={USA}
}

\begin{abstract}
Weak gravitational lensing is one of the few direct methods to map the dark-matter distribution on large scales in the Universe, and to estimate cosmological parameters.
We study a Bayesian inference problem where the data covariance $\mat C$,
estimated from a number $n_{\textrm{s}}$ of numerical simulations, is singular. In a
cosmological context of large-scale
structure observations, the creation of a large number of such $N$-body
simulations is often prohibitively expensive.
Inference based on a likelihood function
often includes a precision matrix, $\Psi = \mat C^{-1}$.
The covariance matrix
corresponding to a $p$-dimensional data vector 
is singular for $p \ge n_{\textrm{s}}$, in which case the precision matrix is unavailable.
We propose the likelihood-free inference method
Approximate Bayesian Computation (ABC) as a solution that circumvents the inversion
of the singular covariance matrix.
We present examples of increasing degree of complexity, culminating in a realistic cosmological scenario of the determination of the weak-gravitational lensing power spectrum for the upcoming European Space Agency satellite Euclid.
While we found the ABC parameter estimate variances to be mildly larger compared to likelihood-based approaches, which are restricted to settings with $p < n_{\textrm{s}}$, we
obtain unbiased parameter estimates with ABC even in extreme cases where
$p / n_{\textrm{s}} \gg 1$.

The code has been made publicly available\footnote{\url{https://github.com/emilleishida/CorrMatrix_ABC}} to ensure the reproducibility of the results.
\ghlink
\end{abstract}


\end{frontmatter}

%%%%%%%%%%%%%%%%% BODY OF PAPER %%%%%%%%%%%%%%%%%%

% from aa.cls
\def\aj{AJ }%
          % Astronomical Journal
\def\araa{ARA\&A}%
          % Annual Revie  of Astron and Astrophys
\def\apj{ApJ }%
          % Astrophysical Journal
\def\apjl{ApJ }%
          % Astrophysical Journal, Letters
\def\apjs{ApJS}%
          % Astrophysical Journal, Supplement
\def\ao{Appl.~Opt.}%
          % Applied Optics
\def\apss{Ap\&SS}%
          % Astrophysics and Space Science
\def\aap{A\&A }%
          % Astronomy and Astrophysics
\def\aapr{A\&A~Rev.}%
          % Astronomy and Astrophysics Reviews
\def\aaps{A\&AS}%
          % Astronomy and Astrophysics, Supplement
\def\azh{AZh}%
          % Astronomicheskii Zhurnal
\def\baas{BAAS}%
          % Bulletin of the AAS
\def\jcap{JCAP}%
	  % Journal of Cosmology and Astroparticle Physics
\def\jrasc{JRASC}%
          % Journal of the RAS of Canada
\def\memras{MmRAS}%
          % Memoirs of the RAS
\def\mnras{MNRAS }%
          % Monthly Notices of the RAS
\def\pra{Phys.~Rev.~A}%
          % Physical Review A: General Physics
\def\prb{Phys.~Rev.~B}%
          % Physical Review B: Solid State
\def\prc{Phys.~Rev.~C}%
          % Physical Review C
\def\prd{Phys.~Rev.~D}%
          % Physical Review D
\def\pre{Phys.~Rev.~E}%
          % Physical Review E
\def\prl{Phys.~Rev.~Lett.}%
          % Physical Review Letters
\def\pasp{PASP}%
          % Publications of the ASP
\def\pasj{PASJ}%
          % Publications of the ASJ
\def\qjras{QJRAS}%
          % Quarterly Journal of the RAS
\def\ropp{Rep. Prog. Phys.}%
	  % Reports on Progress in Physics
\def\skytel{S\&T}%
          % Sky and Telescope
\def\solphys{Sol.~Phys.}%
\def\sovast{Soviet~Ast.}%
          % Soviet Astronomy
\def\ssr{Space~Sci.~Rev.}%
          % Space Science Reviews
\def\zap{ZAp}%
          % Zeitschrift fuer Astrophysik
\def\nat{Nature}%
          % Nature
\def\iaucirc{IAU~Circ.}%
          % IAU Cirulars
\def\aplett{Astrophys.~Lett.}%
          % Astrophysics Letters
\def\apspr{Astrophys.~Space~Phys.~Res.}%
          % Astrophysics Space Physics Research
\def\bain{Bull.~Astron.~Inst.~Netherlands}%
          % Bulletin Astronomical Institute of the Netherlands
\def\fcp{Fund.~Cosmic~Phys.}%
          % Fundamental Cosmic Physics
\def\gca{Geochim.~Cosmochim.~Acta}%
          % Geochimica Cosmochimica Acta
\def\grl{Geophys.~Res.~Lett.}%
          % Geophysics Research Letters
\def\jcp{J.~Chem.~Phys.}%
          % Journal of Chemical Physics
\def\jgr{J.~Geophys.~Res.}%
          % Journal of Geophysics Research
\def\jqsrt{J.~Quant.~Spec.~Radiat.~Transf.}%
          % Journal of Quantitiative Spectroscopy and Radiative Trasfer
\def\memsai{Mem.~Soc.~Astron.~Italiana}%
          % Mem. Societa Astronomica Italiana
\def\nphysa{Nucl.~Phys.~A}%
          % Nuclear Physics A
\def\physrep{Phys.~Rep.}%
          % Physics Reports
\def\physscr{Phys.~Scr}%
          % Physica Scripta
\def\planss{Planet.~Space~Sci.}%
          % Planetary Space Science
\def\procspie{Proc.~SPIE}%
          % Proceedings of the SPIE
\let\astap=\aap
\let\apjlett=\apjl
\let\apjsupp=\apjs
\let\applopt=\ao

\section{Introduction}

Matter on large scales in our Universe is distributed in structures such as walls, filaments, and halos that form the so-called \textit{cosmic web}  \cite[e.g.,][]{1970A&A.....5...84Z,1978IAUS...79..241J}, and most of the matter is composed of dark matter.
A powerful probe to measure statistical properties of the
cosmic (dark-)matter density is weak gravitational lensing, which is the phenomenon of light rays being
deflected by the tidal gravitational fields\footnote{gravity gradients induced
by inhomogeneous matter density distributions} of the large-scale structure
(LSS). These deflections result in the images of observed galaxies being distorted coherently, which is referred to as \textit{cosmic
shear}. The measurement of galaxy shapes and their correlations allow us to infer certain statistical and time-varying properties
of the LSS, and the geometry of the Universe.

Cosmological inference attempts to determine the parameters of the
cosmological model that describes the LSS and geometry of the Universe at
different epochs using observed galaxies at a range of distances from us.
Weak lensing is most sensitive to two cosmological parameters: 
(i) the matter density parameter $\Omega_\textrm{m}$, which is the present-day ratio of the matter density, $\rho_\mathrm{m}$, and critical density, $\rho_\mathrm{crit} = 3 H_0^2 / (8 \pi G)$, where $H_0 = 100 \, h $ km s$^{-1}$ Mpc$^{-1}$ is the Hubble constant and $G$ is Newton's gravitational constant; (ii) $\sigma_8$,
the standard deviation of density fluctuations in spheres of $8 h^{-1}$ Mpc radius. The parameter $\sigma_8$ is also the normalisation of the density power spectrum, which we introduce later.

One of the main weak-lensing observables is the angular power spectrum of
cosmic shear, $C_\ell$, which measures shear correlations induced by the LSS as a
function of angular Fourier mode on the sky, denoted by $\ell$.  A mode $\ell$ denotes a spatial frequency on the sphere, which is related
to angular distance on the sky $\vartheta$ as $\vartheta = \pi / \ell$.  Due to
the non-linear and non-Gaussian evolution of the cosmic matter density field,
different scales become correlated. For weak-lensing inference of cosmological
parameters, these correlations need to be included to avoid biasing the
estimated parameters and their uncertainties
\citep{2018JCAP...10..053B,2019JCAP...08..021S}, which is accomplished by estimating the covariance
of the lensing observables \citep{1998ApJ...498...26K,SvWKM02}.

Despite the non-Gaussian nature of the observables, a multivariate normal
distribution for the likelihood is a good approximation in many cases
\citep[e.g.,][]{CFHTLenS-3pt,2019MNRAS.485.2956H}. This is because the intrinsically non-Gaussian shear is estimated by measuring galaxy shapes where the corresponding noise contribution is due to the variability of the intrinsic shape of galaxies, which is uncorrelated and Gaussian. This noise is particularly important on small and intermediate scales, where the intrinsic shear field is non-Gaussian. This renders the distribution of observed shear more Gaussian, and biases on cosmological parameter from inference under a multivariate normal likelihood approximation are small \citep{2020MNRAS.499.2977L}.

A multivariate normal distribution depends on the data  covariance matrix $\mat C$. To obtain this
matrix for high-dimensional and correlated data is challenging. Several options
have been pursued for cosmic shear. One option is based on analytical calculation using a model prediction \citep{2013arXiv1306.4684K}. This typically involves higher-order
statistics for which models are not well known. Moreover, astrophysical and
observational systematic effects such as intrinsic galaxy alignment for weak lensing
\citep{iareview1} are difficult to model. Another approach is estimation from the data
\citep{2016MNRAS.456.2662F}, which requires a sufficiently large data volume to
create sub-samples. In addition, estimators such as Jackknife re-sampling are
likely to provide results with an unacceptable bias. A third approach uses numerical
simulations of the data, such as $N$-body simulations \citep{2018MNRAS.481.1337H}.
These are typically very expensive so that only a relatively small number
$n_{\textrm{s}}$ of independent simulations are available for a restricted
number of cosmological models.

An additional difficulty comes from the fact that evaluating the likelihood
function requires the inverse of the covariance $\mat C$ (i.e., the
precision matrix $\mat \Psi$). Since matrix inversion is a highly non-linear
operation, uncertainties in the elements of $\mat C$ are amplified and spread
throughout $\mat \Psi$. This mixing of matrix elements can create undesired
correlations. In a cosmological context this can correlate errors on different scales,
which could be the case if simulations are unreliable or biased
on certain scales. For example, dark-matter-only $N$-body simulations are
biased on small scales due to the lack of baryonic physics in those
simulations \citep{2019OJAp....2E...4C}. 
In this paper we focus on the simulation-based estimation of the covariance matrix. The precision and accuracy of the estimated precision matrix,
$\hat{\mat \Psi}$, depend on the ratio $p / n_{\rm s}$, where $p$ is the
dimension of the data (in our setting, this is the number of $\ell$-modes on
which the cosmic-shear power spectrum is measured), and the number of available
simulations $n_{\rm s}$. The smaller $p / n_{\rm s}$ is,
the more precise the matrix estimation. If $p / n_{\rm s} \ge 1$, the estimated covariance matrix is singular, and the
precision matrix is undefined \citep{olkin1954}. Even if $p / n_{\rm s} < 1$
the estimated $\hat{\mat \Psi}$
suffers from a variance that depends on $p / n_{\rm s}$ and increases the uncertainty of the estimated parameters. To reach a desired high accuracy of the estimated parameters from likelihood inference, the ratio $p / n_{\rm s}$ has
to be much smaller than unity \citep{2013MNRAS.tmp.1312T}.

Current cosmological observations from dedicated experiments or surveys
provide data vectors used for inference with typical sizes of order $p \sim
100$.
For future experiments such as the Vera Rubin Observatory (VRO)
Legacy Survey of Space and Time \citep{2009arXiv0912.0201L} or the ESA space
mission Euclid \citep{Euclid-prep-VII_ISTF_20}, this number can be thousands,
or even tens of thousands when weak-lensing is combined with other cosmological probes. 
Keeping the $p / n_{\rm s}$ ratio low would therefore require a large number of $N$-body simulations of LSS, which includes modeling the mutual
gravitational interaction of $N \sim {\cal O}(10^{9 -11})$ mass points over
cosmic times. Such simulations demand  substantial computation time, memory, and
storage space; hence, producing them in these large numbers for the estimation of $\mat \Psi$ is extremely challenging.

To address these challenges, we develop an approach for inference with weak gravitational lensing using Approximate Bayesian Computation \citep[ABC;
e.g.,][]{BeaumontEtAl2002,Beaumont:2010aa, 2011arXiv1101.0955M} that does not require the estimation of the precision matrix, but only the covariance matrix.
An important advantage of using ABC is that a likelihood function does not need to be specified, but instead a forward model is used to generate realizations that are then compared to the real observations. However, this does not preclude the use of a likelihood-motivated generative model. In the proposed method, the generative model is motivated by a Gaussian likelihood function, but with ABC an inverse covariance matrix is not needed. 
One of the key benefits of this approach is that it allows for the setting where $p / n_{\rm s} \ge 1$. For parameter inference with ABC we do not need
to evaluate the likelihood function, $L$, but instead can produce random samples from the data-generating distribution; in our weak-lensing setting where assuming a multivariate normal distribution is considered reasonable, this only requires an estimate of $\mat C$.

ABC has been explored in previous work for weak lensing for various observables such as weak-lensing peak counts \citep{LK15b,LKS16}, which is the number of maxima
in a (filtered) observed shear map. Peaks are a sensitive cosmological probe of non-Gaussian structures,
since they trace the over-dense, highly non-linear regions in the cosmic
density field. Further 
{ABC}
studies were undertaken for the cosmic-shear power spectrum
\citep{2019arXiv190601018K}, and its Fourier transform, the correlation
function \citep{2019arXiv190811523D}.

While the focus of this paper is weak gravitational lensing, the methods developed are applicable to other settings where bypassing the direct estimation of the precision matrix is desirable.  However, a way to simulate realizations of the data is needed.

In this paper we show empirically that our proposed ABC approach can provide unbiased
parameter estimates and uncertainties that are relatively stable with respect
to the number of simulations used for the covariance matrix estimation. We find that the number of simulations can be as low as $n_\textrm{s} = 2$ and still produce reasonable estimates of $\mat C$.
As noted above, an incorrect precision matrix can bias likelihood-based inferred parameters,
see also \cite{2019JCAP...08..021S}. Such biases are amplified during the matrix inversion, and localised errors
can spread across the entire matrix. Our approach minimizes such undesired effects,
and we can expect a lower sensitivity of the inferred parameters to covariance matrix errors.
We emphasize that we are using a likelihood-motivated forward model for the ABC algorithm; a different generative model could also be considered that does not require a covariance matrix, but may have other costs such as more computationally-intensive data generation.

The article is organised as follows. We provide a brief introduction to
cosmological inference with weak gravitational lensing in
Sect.~\ref{sec:background} to motivate our approach. Section
\ref{sec:ABC_descr} gives an overview of our methodology using ABC. Section
\ref{sec:sim_study} discusses our three simulation studies that demonstrate the
performance and limits of the proposed approach. These include a simple linear
model (Sect.~\ref{sec:toy_example}), a non-linear weak-lensing inspired
function (Sect.~\ref{sec:WL_example}), and a realistic weak-lensing case
(Sect.~\ref{sec:WL_model}). Concluding remarks are presented in
Sect.~\ref{sec:conclusions}.

%%%%%%%%%%%%%%%%%%%%%%%%%%%%%%%%%%%%%%%
\section{Background}
\label{sec:background}

We begin this section by providing additional background on our motivating example of cosmological inference using weak lensing, including a description of the statistical model of the corresponding observable. Then we present some details about covariance and precision matrix estimation, which motivates our use of ABC.

\subsection{Weak gravitational lensing}

In most results from gravitational lensing observations to date, the data are second-order statistics of the weak-lensing cosmic shear field, and thus can be derived from the shear power spectrum $C_\ell$. The power spectrum is a one-dimensional function of angular Fourier scale $\ell$, which is the modulus of the 2D Fourier vector $\vec \ell$ on the sky. 

\subsubsection{The density contrast power spectrum}

Before defining the weak-lensing power spectrum $C_\ell$, several other quantities need to be introduced. The density contrast, $\delta(\vec x, z) = [\rho(\vec x,
z) - \bar \rho(z)] / \bar \rho(z)$, is the scaled matter density
fluctuation $\rho$
around the spatial mean, $\bar \rho$, at redshift $z$, and 3D position $\vec x$.
Redshift is the relative wavelength change between emitted and observed radiation
of an object. In general, due to the expansion of the Universe the receding velocity of a galaxy increases with distance, which results in a larger redshift. Redshift can be used as a proxy for cosmological distance. The position vector $\vec x$ is a comoving coordinate, which remains constant with the expansion of the Universe.
At early times and on large scales, $\left| \delta \right|
\ll 1$, and the evolution of the density can be described by linear Newtonian
perturbation theory \citep[e.g.,][]{pee80,1998ApJ...496..605E}. On small scales,
the non-linear evolution due to gravitational collapse is typically modeled by
phenomenological approaches, e.g.,~parameterized models of halo formation and clustering, some based on numerical simulations
\citep{2012ApJ...761..152T}.

The first moment of the density field vanishes by definition, $\mathbb E \left( \delta \right) = 0$. Its second moment, or two-point correlation function $\xi_\delta$ is defined as
\begin{align}
    \xi_\delta(\vec x_1, \vec x_2)  
    = \mathbb E \left[ \delta(\vec x_1) \delta(\vec x_2) \right]
    = \int \int  \delta(\vec x_1) \delta(\vec x_2) P[\delta(\vec x_1), \delta(\vec x_2)] \mathrm{d} \delta(\vec x_1) \mathrm{d} \delta(\vec x_2),
\label{eq:cf}
\end{align}
where
$P[\delta(\vec x_1), \delta(\vec x_2)]$
is the joint probability distribution function of $\delta(\vec x_1)$ at position $\vec x_1$ and $\delta(\vec x_2)$ at $\vec x_2$.

We assume that on large scales the density field is statistically homogeneous and isotropic. In that case, the two-point correlation function is invariant under translation and rotation such that $\xi_\delta(\vec x_1, \vec x_2) = \xi_\delta(|\vec x_1 - \vec x_2|)$. In addition, the Fourier transform of $\xi_\delta$, known as the 3D power spectrum of the density field $P_\delta$, can be written as
\begin{equation}
P_\delta(k) = \int_{\mathbb R^3} \mathrm{d}^3 y \, \mathrm{e}^{\mathrm{i} \vec k \vec y} \xi_\delta(y).
\label{eq:P_delta}
\end{equation}
Due to statistical isotropy the power spectrum only depends on the modulus of the 3D wave mode, $k = |\vec k|$.

\subsubsection{The weak-lensing power spectrum}

For weak gravitational lensing by galaxies, the basic observable is the
shape of a galaxy, expressed as complex ellipticity $\varepsilon$. Lensing by
the LSS changes the galaxy shape, and imprints a shear
$\gamma$. If a galaxy has an intrinsic ellipticity $\varepsilon^{\rm s}$, to
first order, the following relation holds,
\begin{equation}
  \varepsilon \approx \varepsilon^{\rm s} + \gamma.
\end{equation}
Thus, the lensing shear $\gamma$ can be estimated by the observed ellipticity of galaxies, noting
that the expected value of their intrinsic ellipticity vanishes.

The lensing shear power spectrum is a projection, or integral along the redshift direction of the matter density power spectrum of 
Eq.~\eqref{eq:P_delta}.
The following equation shows the result
for a flat universe where the spatial curvature vanishes, and consequently the total (matter + radiation + cosmological constant) density is equal to the critical density $\rho_\mathrm{crit}$ at all times.
Integrating along comoving distance $\chi$, which 
is related to redshift as
${\rm d} \chi  = - {\rm d} z H^{-1}(z)$,
with the Hubble parameter $H(z) = H_0 \left[ \Omega_\textrm{m} (1 + z)^3 + 1 - \Omega_\textrm{m} \right]^{1/2}$ for a flat universe,
we get
\begin{equation}
  C_\ell = \frac 9 4 \, \Omega_{\rm m}^2 \left( \frac{H_0}{c} \right)^4
  \int_0^{\chi_{\rm lim}} {\rm d} \chi \,
  [1+z(\chi)]^2 q^2(\chi) P_\delta\left(\frac{\ell}{\chi}, \chi \right),
  \label{eq:p_kappa_limber}
\end{equation}
where $c$ is the speed of light.
This equation holds under some approximations that are very good in most
practical cases \citep{1953ApJ...117..134L,2016arXiv161104954K,KH17}.
The lensing efficiency $q$ is given as an integral over the normalised galaxy
number count ${\rm d} \chi \, n(\chi)$ and geometrical factors, as
\begin{equation}
  q(\chi) = \int\limits_\chi^{\chi_{\rm lim}} {\rm d} \chi^\prime \, n(\chi^\prime)
  \frac{\chi^\prime - \chi}{\chi^\prime},
  \label{eq:lens_efficiency}
\end{equation}

\begin{figure}
    \centering
    \includegraphics[width=0.715\textwidth]{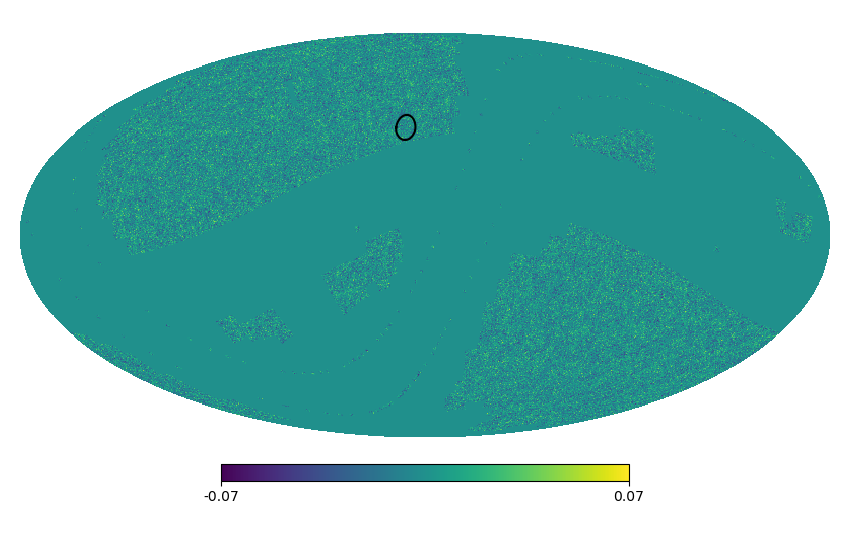}
    \raisebox{5em}{%
      \includegraphics[width=0.25\textwidth]{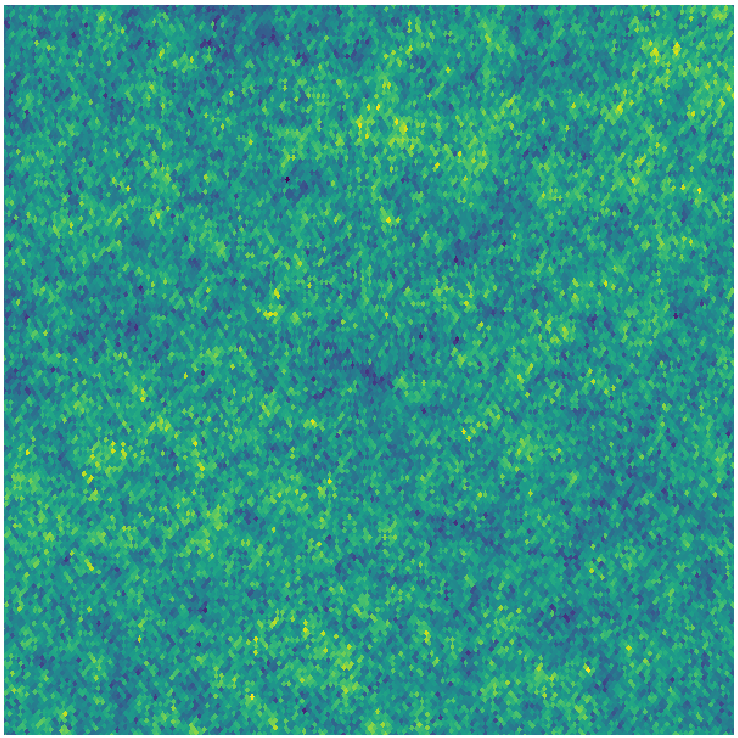}
    }

    \caption{A map of the convergence sampled from a log-normal distribution with power spectrum Eq.~\eqref{eq:p_kappa_limber} and parameters as described in Sect.~\ref{sec:WL_model}.
    The colors encode the values of the weak-lensing convergence as indicated by the color bar.
    \textit{Left panel:} Full sky map in Mollweide projection, ecliptic coordinates, showing the Euclid footprint, see \citet{2021arXiv210801201S} for more details.
    The large bands are
    the Ecliptic and Galactic plane, which are masked due to dust extinction and zodiacal light.
    \textit{Right plot:} Zoom-in of a $10 \times 10$ deg$^2$ area, marked by the circle in the left image.
    }
    \label{fig:lensing_map}
\end{figure}

The lensing power spectrum of Eq.~\eqref{eq:p_kappa_limber} is the most fundamental
observable for cosmological inference as it depends on cosmological parameters
in several ways. It is a linear projection of the 3D matter power spectrum
$P_\delta$, which encodes the matter distribution and its evolution over
redshift. The projection kernel is a function of the comoving distance
$\chi$, which depends on the matter content and the geometry of the
Universe. In addition, it is directly affected by two cosmological parameters:  the matter density $\Omega_\textrm{m}$ and the Hubble constant
$H_0$.

The $C_\ell$ describes the power in the shear field $\gamma$ as a function of projected
Fourier distance on the sky $\ell$. We write in the flat-sky approximation
\begin{equation}
   \left\langle \tilde \gamma(\vec \ell) \tilde \gamma^\ast(\vec \ell^\prime) \right\rangle
    = \delta_\textrm{D}^{(2)}(\vec \ell - \vec \ell^\prime) \, C_\ell,
    \label{eq:gammagamma}
\end{equation}
where $\tilde \gamma$ denotes the Fourier transform of $\gamma$, and the superscript `$^\ast$' denotes the complex conjugation. Due to statistical homogeneity, different Fourier modes of the field $\gamma$ are uncorrelated, which is expressed by the 2D Dirac delta distribution $\delta_\textrm{D}^{(2)}$.
On large scales, $\ell \lesssim 100$, the
shear field is very close to Gaussian, and different Fourier modes of the power spectrum
are uncorrelated. The $C_\ell$ has a maximum at $\ell$ of a few $10$, which
corresponds to the size of the horizon at the time of matter-density
equality. This is defined as the time in the early Universe when the matter
density equals the radiation density, and matter subsequently became dominant, at a
redshift of $z_{\rm eq} \approx 31,000 \, \Omega_\mathrm{m} h^2$, in the early Universe. The non-linear regime starts at $\ell
\gtrsim 1,000$, marked by an increase of power, and the convergence field becomes
non-Gaussian resulting in the power spectrum becoming correlated across different $\ell$-modes.
An example of the lensing power spectrum will be presented in Sect.~\ref{sec:exp2_data_gen}.

\subsubsection{The lensing power-spectrum covariance}
\label{sec:WL_cov}

For the observed lensing power spectrum, $C_{\ell, {\rm obs}}$, we need to account
for shot noise, which is due to intrinsic galaxy ellipticities having a
non-vanishing dispersion $\sigma_\varepsilon$ of typically $0.3$ - $0.4$.
We write
\begin{equation}
C_{\ell, {\rm obs}} = C_\ell + \frac{\sigma_\varepsilon^2}{\bar n},
\label{eq:C_obs}
\end{equation}
In the linear regime for the density contrast, the lensing power spectrum can
be approximated by a normal distribution with uncorrelated Fourier
modes. The variance in this approximation is
\begin{equation}
\mathbb E \left( C_{\ell, \OBS}^2 \right) - \mathbb E \left( C_{\ell, \OBS} \right)^2 = \frac{1}{f_{\textrm{sky}} (2 \ell + 1)}
C_{\ell, \OBS}^2,
\label{eq:cov_WL_G}
\end{equation}
where $f_{\textrm{sky}}$ is the fraction of the observed sky area
\citep{1992ApJ...388..272K}.
This diagonal covariance matrix will be referred to as the \textit{Gaussian covariance}.

This Gaussian covariance is appropriate in the regime where the Fourier modes of the lensing field are independent. This is no longer true on small scales, where the non-linear evolution of the density field leads to mode coupling. 
Further, there is a coupling of non-linear, small scales to long-wavelength modes  that can be larger than the observed area. This is due to the finite observed volume, which is not representative of the ensemble, large-scale density fluctuations. These fluctuations induce additional power on small scales, which create the coupling between small and large scales.
These mode couplings mean that different modes of the field $\gamma$ are no longer independent, and the distribution of $\gamma$ becomes non-Gaussian. This implies that the covariance of the convergence power spectrum is no longer diagonal. Note that the Fourier modes of $\gamma$ itself are still uncorrelated, since Eq.~\eqref{eq:gammagamma} is based only on the homogeneity of $\gamma$.
The small-to-long wavelength mode couplings lead to an additional lensing covariance, which is the dominant non-Gaussian contribution \citep{2018JCAP...10..053B}. This term is called \textit{super-survey covariance} (SSC).
\citet{2018JCAP...10..053B} showed that the total weak-lensing
power-spectrum covariance is well represented by the Gaussian covariance plus the SSC term, which we refer to as the \textit{Gaussian plus SSC covariance}.

Eq.~\eqref{eq:p_kappa_limber} is the power spectrum of not only the shear, but of a scalar field called the convergence. This field can be obtained from the shear field, and is more directly related to the density contrast.
Fig.~\ref{fig:lensing_map} shows a realisation of a convergence field, drawn
from a log-normal distribution with powe~r spectrum Eq.~\eqref{eq:p_kappa_limber}.
The cosmological parameters and redshift distribution that are used are
described in Sect.~\ref{sec:WL_model}. To the full-sky convergence map the
footprint mask of the Euclid survey is applied, resulting in a remaining observed area of $f_\mathrm{sky} = 0.3636$ corresponding to $15,000$
deg$^2$.
Euclid \citep{2011arXiv1110.3193L} is a European space experiment, to be launched in early 2023. Euclid will observe $1.5$ billion galaxies in optical and infrared wavelengths. Weak gravitational lensing is one of the two main cosmological observables for Euclid.
The Euclid mask cuts out unobserved regions, with the main areas being the Galactic and the Ecliptic plane. These regions on the
sky are heavily influenced by dust in the Milky Way disk, and zodiacal
light (reflection of sun light by dust particles), respectively. These effects adversely
impact the number of observable background galaxies, their flux and their
measured shapes.

\subsubsection{The lensing likelihood function}

We suppose that the observed power spectrum $C_\ell$ is measured at $p$ discrete Fourier scales $\ell_i, i=1 \ldots p$.
The data vector $\vec y = \{ C_{\ell_i} \}$ is assumed to follow a $p$-dimensional multivariate
normal distribution, $\vec y \sim {\cal N}_p(\vec m(\vec \theta), \mat C)$, with mean
$\vec m(\vec \theta) \in \mathbb R^p$ and $p \times p$ covariance matrix $\mat
C$.
The mean depends on the parameter vector
$\vec \theta \in \mathbb R^{n_\theta}$.
The likelihood function $L$ is the usual multivariate normal likelihood given in log form by
\begin{align}
	-\ln L(\vec y| \vec \theta) & 
    = \, {\rm const} + \frac 1 2 \ln \tr
    	\, \mat C + \frac 1 2 \chi^2(\vec y | \vec \theta); \nonumber \\ 
  \chi^2(\vec y | \vec \theta)  & = \left[ \vec y - \vec m(\vec \theta) \right]^{\rm t} \mat \Psi
    	\left[ \vec y - \vec m(\vec \theta) \right] ,
    \label{eq:mvnorm}
\end{align}
which depends on the precision matrix, $\mat \Psi = \mat C^{-1}$.
The covariance and precision matrices are considered independent of $\vec\theta$.

\subsection{Estimators of the covariance and precision matrices}

A summary of estimators for the covariance matrix ($\hat{\mat C}$), and a biased ($\hat{\mat \Psi}_\ast$) and unbiased ($\hat{\mat \Psi}$) estimator of the precision matrix are displayed in Table~\ref{tab:estimators}.  Along with the estimators, their first and second moments are also included.  
Note that throughout we use the notation $\vec z^\mathrm{t}$ to indicate the transpose of $\vec z$, and given some matrix $\mat Z$, $Z_{ij}$ represents the element in the $i$th row and $j$th column of $\mat Z$.  
In Table~\ref{tab:estimators}, the estimators consider simulated realizations of the data vector, $\vec y_j \in \mathbb R^p$, with $j = 1, \ldots, n_{\rm s}$.  The sample mean vector is $\vec{\bar y}\in \mathbb R^p$, where $\vec{\bar y} = n_{\rm s}^{-1} \sum_{j=1}^{n_{\rm s}} \vec y_j$.

For $\vec y \sim {\cal N}(\vec m(\vec \theta), \mat C)$, the covariance estimator is $p$-Wishart-distributed, $\nu\hat{\mat C} \sim {\cal W}_p(\mat C,\nu)$, with degrees of freedom $\nu = n_{\rm s} - 1$. 
The Wishart distribution is well-defined for $\nu > p - 1$, for which $\hat{\mat C}$ is invertible. 
For $\nu \le p - 1$, $\hat{\mat C}$ is singular,
and $\hat{\mat C}$ follows a singular or anti-Wishart distribution \citep{MKM79}. 

A biased estimator of the precision matrix is $\hat{\mat \Psi}_\ast$ \citep{Siskind72}, as displayed in Table~\ref{tab:estimators}.
This estimator follows an inverse-Wishart distribution, $\nu \hat{\mat \Psi}_\ast \sim {\cal W}_p^{-1}\left(\mat \Psi, \nu\right)$, with degrees of freedom $\nu$. A debiased estimator can be found as $\hat{\mat \Psi} =\left(\nu - p - 1\right)\nu^{-1}\hat{\mat \Psi}_\ast$.
This result has been routinely used to debias the estimated covariance from simulations in weak gravitational lensing \citep[e.g.,][]{HSS07,2017MNRAS.465.2033J,2017arXiv170609359K},  and cosmology in general \citep[e.g.,][]{PhysRevD.79.083012,2014MNRAS.439.2531P,2018MNRAS.481.1337H}.

\begin{table}[]
    \centering
    \begin{tabular}{|c|c|c|c|}
    \hline
Symbol & Description & Form \\
    \hline 
    %%%----------------------------------------------------
    $\hat{\mat C}$ & Covariance  & $\hat{\mat C} = \frac 1 {n_{\rm s} - 1} \sum_{j=1}^{n_{\rm s}}
    	\left({\vec{y}^j} - \vec{\bar y} \right)
        \left({\vec{y}^j} - \vec{\bar y} \right)^{\rm t}$  \\
        &  matrix & $\mathbb E\hat{\mat C} = \nu \, \mat S = \mat C$  \\
        & & $ \mathbb E(\hat C_{ij} - C_{ij})^2 = \nu^{-1} \left( C_{ij}^2 + C_{ii} C_{jj} \right)$, \citep{doi:10.1080/02331888808802132}  \\
    \hline 
    %%%----------------------------------------------------
    $\hat{\mat \Psi}_\ast$ & Precision  & $\hat{\mat \Psi}_\ast = \hat{\mat C}^{-1}$  \\ 
    & matrix & $\mathbb E(\hat{\mat \Psi}_\ast) = (\nu - p - 1)^{-1} \nu \, \mat C^{-1}$  \\
    & & for $\nu > p - 1$,  \citep{doi:10.1080/02331888808802132} \\
    & & $\mathbb E( \hat \Psi_{\ast ij} - \Psi_{\ast ij} )^2 = \frac{\nu^2 \left[ \left( \nu - p + 1 \right) \Psi_{ij}^2 + (\nu - p - 1) \Psi_{ii} \Psi_{jj}
    \right]}{(\nu - p) (\nu - p - 1)^2 (\nu - p - 3)}$,  \\
    & & for $\nu > p + 3$, \citep{CF11} \\
        \hline 
    %%%----------------------------------------------------
    $\hat{\mat \Psi}$ & Precision & $\hat{\mat \Psi} = \alpha \hat{\mat \Psi}_\ast$  \\
    & matrix & $\alpha = \frac{\nu - p - 1}{\nu} = \frac{n_{\textrm{s}} - p - 2}{n_{\textrm{s}} - 1}$  \\
    & & $\mathbb E(\hat{\mat \Psi}) = \mat C^{-1}$  \\
    & & $\mathbb E( \hat \Psi_{ij} - \Psi_{ij} )^2 = \alpha^2 \mathbb E( \hat \Psi_{\ast ij} - \Psi_{\ast ij} )^2$  \\
        \hline 
    %%%----------------------------------------------------
    \end{tabular}
    \caption{Estimators and two moments for covariance and precision matrices.}
    \label{tab:estimators}
\end{table}

\begin{SCfigure}
    \resizebox{0.6\columnwidth}{!}{
    \includegraphics{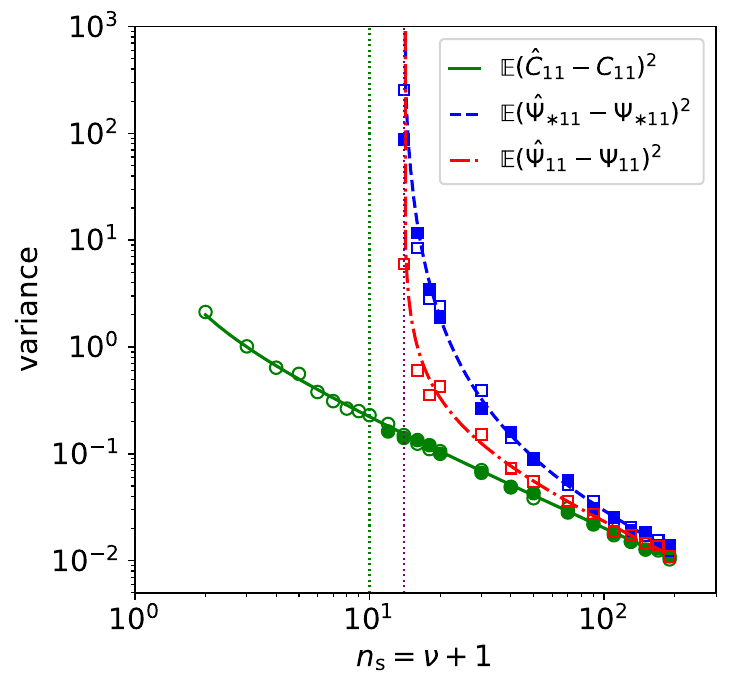}
     }

     \caption{Variance of the Wishart (${\cal W}_{10}$, green solid line/circles)
     and
     inverse Wishart (${\cal W}_{10}^{-1}$, dashed blue line/squares) distributions, as function
     of $\nu+1$. The dash-dotted line
     and squares show the debiased variance of ${\cal W}_{10}^{-1}$.
     Values are $(1,1)$ matrix elements, lines analytical expressions, filled symbols
     samples from ${\cal W}_{10}$ and ${\cal W}_{10}^{-1}$.
     Open symbols use samples from ${
     \cal N}$. The vertical lines 
     indicate the
     limit of ${\cal W}_{10}$ (${\cal W}_{10}^{-1}$), 
     at $\nu = p - 1$ ($\nu
     = p + 3$).
     \nblink{cov_matrix_definition/wishart}
     }

     \label{fig:W_variance}
 \end{SCfigure}

To show how the analytical expressions for the variances of the estimators compare to 
(i) the estimated variances computed based on random draws from the sampling distributions of the estimators and (ii) the estimated variances computed from random draws of the data vector from a multivariate normal, we carried out a comparison by doing the following. 
Consider a $10 \times 10$ identity covariance matrix $\mat C = \mat I_{10}$.  We plot the variance of the $(1,1)$-element of $\hat{\mat C}$ (green),
$\hat{\mat \Psi}_\ast$ (blue), and $\hat{\mat \Psi}$ (red) in
Fig.~\ref{fig:W_variance}. The analytical expressions, shown as lines, are
well-reproduced when sampling from ${\cal W}_{10}(\mat C, n_{\rm s}-1)$ and ${\cal W}_{10}^{-1}(\mat C^{-1}, n_{\rm s}-1)$ (filled symbols).
We also sample a vector $\vec y$ from a
multivariate normal distribution, calculate $\hat{\mat C}$, $\hat{\mat \Psi}_\ast$, and $\hat{\mat \Psi}$ using
the estimators from Table~\ref{tab:estimators}, and the
variance of the estimators from the $\nu + 1$ simulations (open symbols).
The key point is that we can estimate mean and variance of
the covariance estimator $\hat{\mat C}$ for $\nu \le p - 1$ by simulating realizations from the appropriate multivariate normal distribution, even in cases where the inverse covariance is no longer defined (green open circles).
We make extensive use of this feature in the proposed methodology.

\section{Approximate Bayesian Computation (ABC)}
\label{sec:ABC_descr}

Some of the problems that plague traditional likelihood-based inference methods can be circumvented with Approximate Bayesian Computation \citep[ABC;][]{TavareEtAl1997,PitchardEtAl1999,BeaumontEtAl2002,Beaumont:2010aa}.
Rather than specifying a likelihood function, ABC works by drawing samples from a simulation model, $\vec y\Sim \sim M(\vec \theta)$.  The basic ABC algorithm \citep{TavareEtAl1997,PitchardEtAl1999} samples values of $\vec \theta^{*}$ from some user-specified prior(s), $\pi(\vec \theta)$, generating a simulated dataset $\vec y\Sim \sim M(\vec \theta^{*})$, and then comparing $\vec y\Sim$ to the observed dataset, $\vec y\Obs$.  If $\vec y\Sim$ and $\vec y\Obs$ are ``close enough'' (discussed more specifically below), then $\vec \theta^{*}$ is retained and contributes to the particle approximation of the posterior distribution of $\vec \theta$.

For the simulation model, the likelihood function need not to be known
explicitly. For example, the simulation model could, in theory, be derived from
$N$-body simulations if they were computationally feasible for use in an ABC algorithm (which they are not).  Since $N$-body
simulations are too computationally intensive, fast approximate simulations
could be used instead. The simulations generate realizations of observables
through a complex forward process that mimics the gravitational interactions in
an expanding Universe.

ABC can also be carried out by replacing a complex likelihood-free simulation
model with sampling from an explicitly chosen distribution (from which a
likelihood function $L$ would be based). The advantage in this case over
MCMC is that $L$ does not need to be
evaluated. In particular, if the chosen distribution is a multivariate normal, the precision matrix is not required which is important in our setting.

We compare ABC-based parameter estimates and their uncertainties
to the corresponding Fisher matrix-based and MCMC-based quantities. We assume
that we have simulated $n_{\rm s}$ independent multivariate normal random vectors from which we compute an
estimate of the covariance matrix using $\hat{\mat C}$ from Table~\ref{tab:estimators}. The Fisher
matrix and MCMC then require the evaluation of
a likelihood function together with an estimate of the precision matrix.
For ABC we need only to generate data vectors from a simulation model,
which do not depend on $\hat{\mat\Psi}$, but only on $\hat{\mat C}$.
We demonstrate that inference using ABC is possible when $\nu \le p + 1$, even for $p / \nu > 100$, 
and the results do not appear to depend on $\nu$.

\subsection{ABC algorithm}
\label{sec:ABC_algo}

The basic ABC algorithm discussed above can be computationally inefficient so a
number of extensions of this algorithm have been proposed (e.g.,
\citealt{SissonEtAl2007,BeaumontEtAl2009, DelMoralEtAl2011,BonassiWest2004}).
In this work, we use the
\texttt{CosmoABC}\footnote{https://github.com/COINtoolbox/cosmoabc}
implementation of the ABC-Population Monte Carlo (ABC-PMC) algorithm
\citep{BeaumontEtAl2009} described in \cite{2015A&C....13....1I}. 
%The details
%of the algorithm are displayed in Algorithm~\ref{alg:abc}, and discussed next.

ABC-PMC starts with the basic ABC algorithm in iteration $t=1$ where samples of
a $n_\theta$-dimensional parameter vector $\vec \theta$ are proposed from the
prior distribution, $\pi(\vec \theta)$. Simulated data, $\vec y\Sim$, are
generated from the simulation model using a proposed $\vec \theta^*$, and are
then compared to the observations, $\vec y\Obs$, via a distance function
$D(\cdot, \cdot)$.  Often the distance function uses summaries $\vec s(\cdot)$ of the
simulated and observations rather than the data themselves. The distance
functions developed for our setting are discussed in Sect.~\ref{sec:s_d}. If
$D[\vec s(\vec y\Sim), \vec s(\vec y\Obs)] \leq \epsilon_t$, where $\epsilon_t$
is a tolerance value for iteration $t$, then the proposed $\vec \theta^*$ is
retained; otherwise it is discarded.  This is repeated until $M$ values are
accepted, $\{\vec \theta^{(j)}\}_{j=1}^M$.

In subsequent iterations $t = 2, \ldots, T$, rather than drawing directly from $\pi(\vec
\theta)$, the previous iteration's ABC posterior is used where the selected
values are moved according to a user-specified kernel, $K(\vec \theta,
\cdot)$ (in our case, a Gaussian kernel).  Importance weights are
defined to account for this change in proposal distribution after the initial use of the  prior in step $t=1$, and the resulting ABC
posterior distribution is based on the particle system $\{\vec \theta_t^{(j)},
W_t^{(j)}\}_{j = 1}^M$ where $W_t^{(j)}$ are the importance weights such that
$\sum_{j = 1}^M W_t^{(j)} = 1$. 
%(See Algorithm~\ref{alg:abc} for the form of
%the importance weights.)
We update the distance threshold $\epsilon_{t+1}$ of iteration $t+1$ as the $q$ percentile of the distances in iteration $t$, with $q=0.75$.

For the stopping criterion, at each iteration the number of accepted particles divided by the number of proposed particles, $\delta_t$, is computed.
If $\delta_t < \delta_0$, where $\delta_0$ is a user-specified threshold, the sampling is stopped, which occurs when the number of draws exceeds the number of particles by $1/\delta_0$. We refer the reader to \citet{2015A&C....13....1I} for a detailed description of the algorithm. 

\subsection{Simulation set-up}
\label{sec:ABC_set_up}

\subsubsection{Data generation}
\label{sec:set_up_data_gen}

For all our experiments, we simulate observation  $\vec d\Obs = (\vec x\Obs, \vec y\Obs)$,
with $\vec x\Obs = ({x\Obs}_{, 1}, \ldots,$ ${x\Obs}_{, p})$ and
$\vec y\Obs = ({y\Obs}_{, 1}, \ldots, {y\Obs}_{, p})$, as follows. We fix the abscissa vector
$\vec x\Obs$,
and the $p \times p$ covariance matrix $\mat C$.
Then we generate $n_\textrm{r}$ independent draws of the observed ordinate vectors $\vec y^r\Obs$, $r= 1 \ldots n_\textrm{r}$ as multivariate
normal random variable, $\vec y^r\Obs \sim {\cal N}(\vec m(\vec \theta_0), \mat C)$. The
mean is the model prediction for a given fixed model parameter vector $\vec \theta_0$.
Next, for each $r$ and given a simulation sample size $n_\textrm{s} = \nu - 1$ we compute the sample covariance $\hat{\mat C}$. If
$p \le \nu$, we can sample directly from a Wishart distribution, $\hat{\mat C} \sim {\cal W}_p(\mat C, \nu)$.
For $p > \nu$, we instead generate $n_{\rm s}$ multivariate
normal random variables $\vec y^j \sim {\cal N}(\vec y^r, \mat C)$, and compute $\hat{\mat C}$
according to Table \ref{tab:estimators}.

\subsubsection{ABC sampling details}

To create the initial ABC set of particles we draw $2 M$ values from the prior, and retain the $M$ particles that result in the lowest distance. Each subsequent ABC-PMC iteration uses $M$ particles. 
For each simulated observation corresponding to a proposed parameter vector $\vec \theta^*$,
we create a model prediction $\vec y\Sim(\vec \theta^*) \sim {\cal N}(\vec m(\vec \theta^*), \hat{\mat C})$, where $\hat{\mat C}$ is the sample covariance matrix generated earlier (which is singular for $p < \nu$).

\subsection{Summary statistics and distances}
\label{sec:s_d}

The cases discussed in this work simulate one-dimensional functions $f:
\mathbb R \rightarrow \mathbb R$, mimicking the weak-lensing power spectrum $C_\ell$ of Eq.~\eqref{eq:p_kappa_limber}. The data vectors are composed of joint abscissa and ordinate vectors, $\vec d\Z = (\vec
x\Z, \vec y\Z)$, Z $\in$ \{obs, sim\}, where the two identifiers stand for observation and simulation, respectively.

In the following we introduce the distance functions used in this work, that do not depend on the estimated precision matrix. We propose simple distance functions that ignore any correlation between data points, and also a new distance function that accounts for correlation without depending on $\hat\Psi$. 

\subsubsection{Parameter-based distance function}

For a function $f$ which can be described by a parameter vector $\vec \theta \in \mathbb R^{n_\theta}$, a distance function can be constructed based on parameter estimates as follows.
For a given set of simulated data vectors $\vec d\Sim = (\vec x\Sim,
\vec y\Sim)$, the function $f$ is fitted with $\hat{\vec \theta}\Sim$, which is obtained from an ordinary least squares regression. An analogous fit is
performed on the observed data to yield the best-fit parameter $\hat{\vec \theta}\Obs$.
For this case, the summary statistic depends on the best-fit parameters,
\begin{equation}
  \vec s_{\rm param}(\vec d) = \hat{\vec \theta},
  \label{eq:s_param}
\end{equation}
corresponding to a compression of the data into $n_\theta$ values.
Then, the \textit{parameter} distance $D_{\rm param}$ is
\begin{align}
  D_{\rm param}&[\vec s_{\rm param}(\vec d\Obs), \vec s_{\rm param}(\vec d\Sim)]
  = \sqrt{ \sum_{i = 1}^{n_\theta} \left[ {\theta\Obs^\ast}_{, i} - {\theta\Sim^\ast}_{, i} \right]^2}.
    \label{eq:dist_param}
\end{align}

\subsubsection{Covariance-based distance function}

A more general summary statistic $\vec s$, which uses the full data vector, is
\begin{equation}
  \vec s(\vec d) = \vec y.
  \label{eq:summ_data}
\end{equation}
A general distance function is the \textit{Mahalanobis} distance $D_{\mat \Psi}$, given as
\begin{equation}
  D_{\mat \Psi}[\vec s(\vec d\Obs), \vec s(\vec d\Sim)]
  = \sqrt{\Delta \vec y^{\rm t} {\mat \Psi} \Delta \vec y},
  \label{eq:dist_data_invcov}
\end{equation}
where
\begin{equation}
\Delta \vec y = \vec y\Obs - \vec y\Sim
\label{eq:Delta_y}
\end{equation}
This distance can however not be used
if the true precision matrix is not known and the estimated covariance matrix
is singular. We therefore
have to find an alternative distance. A distance derived from Eq.~\eqref{eq:dist_data_invcov} can be obtained by replacing the precision matrix by a diagonal matrix
with the reciprocal elements of the estimated covariance matrix $\hat C$ on the diagonal. This results in the \textit{inverse-variance} distance function
\begin{align}
  D_{\hat C}[\vec s(\vec d\Obs), \vec s(\vec d\Sim)]
  = \sqrt{\sum_i \left( \frac{\Delta y_i}{\hat C_{ii}} \right)^2 },
  \label{eq:dist_data_diag}
\end{align}
The diagonal elements $\hat C_{ii}$ are non-zero even in the case where
$\hat{\mat C}$ is singular.

\subsubsection{Autocorrelation-based distance function}

Next, we introduce a new distance that accounts for the correlation of data points
but does not require the inversion of the covariance matrix. 
Note that with this distance the covariance matrix is only used to generate the simulated data.
This distance is
based on the autocorrelation function (acf) of the data $\vec y\Obs$, which is a function of lag $t$.
We define the unnormalised acf as
\begin{equation}
  \xi_{\vec y}^{\textrm{u}}(t) =
    \frac 1 {(p - t)} \sum_{i=1}^{p-t}
      \left( y_i - \bar y^{(1,p-t)} \right) \, \left( y_{i + t} - \bar y^{(t+1,p)} \right) ; \quad\quad t=0 \ldots p-1,
      \label{eq:xiu}
\end{equation}
where we subtract from each shifted data vector the corresponding mean,
\begin{equation}
  \bar y^{(m, n)} = \frac 1{(m - n)} \sum_{i=m}^n y_i .
\end{equation}
The function $\xi_{\vec y}^{\textrm{u}}(t)$ is normalised such that its value is unity at $t=0$, which gives the acf as the following
\begin{equation}
  \xi_t(\vec y) = 
    \frac{\xi_t^{\textrm{u}}(\vec y)}{\xi_0^{\textrm{u}}(\vec y)} .
  \label{eq:xi}
\end{equation}
The acf quantifies the mean autocorrelation of the data between
two entries with indices separated by $t$, in the case of equidistant abscissa; this
difference is $\Delta x = x_{i+t} - x_i$ for all $i$. The
autocorrelation is used in the distance function to penalize data points that are strongly
correlated with others by increasing the overall distance.  The \textit{acf} distance is defined as
\begin{align}
D_{\rm acf}&[\vec s(\vec d\Obs), \vec s(\vec d\Sim)]
  = \left| \sum_{i, j=1}^{{n_\textrm{d}}}
      \Delta y_i \, \xi_{|i-j|} \, \Delta y_j
  \right| .
  \label{eq:d_acf}
\end{align}
This distance accounts for the correlation between data points without depending on
the precision matrix. For a given difference between data points indices $t=|i-j|$,
the acf $\xi_{|i-j|}$ contributes to the distance as a weight, and represents the average correlation corresponding to that $t$.
Summands in Eq.~\eqref{eq:d_acf} with $i=j$ contribute maximally to the distance, with unit weight
$\xi_0 = 1$. Off-diagonal terms with $i \ne j$ that are not correlated do not
contribute significantly to the distance; the intuition is that the distance should not be influenced by the difference between uncorrelated data points. 
On the other hand, two
correlated data points with $0 < \xi_{|i-j|} \le 1$ contribute to the overall distance, by penalising models that predict $y_{\Sim, i}$ to be very different from
$y_{\Obs, j}$.

\section{Simulation Study}
\label{sec:sim_study}

In this section we present three simulation experiments to evaluate the performance of the proposed ABC method. We investigate the uncertainty in the covariance 
matrix estimation, and its propagation to the parameter estimates, their standard
errors, and the uncertainty in the parameter estimates' errors (i.e., the uncertainty in the uncertainty).
All three experiments simulate one-dimensional functions, where the abscissa values are drawn from a multivariate normal distribution with a given covariance matrix. 
A flat prior is used in each of the $n_\theta = 2$ dimensions of $\vec\theta$.
Next, we describe different methods for comparison with the proposed ABC approach.

%%%------------------------------- COMPARISON METHODS
\subsection{Comparison methods}
\label{sec:mcmc}

The performance of the proposed ABC method is compared to a Fisher information matrix approach where the Fisher matrix is computed for a multivariate normal distribution.
The Fisher matrix is computed at the (true) input values for the parameters.
For the first example, a Metropolis-Hastings Monte-Carlo Markov Chain
(MCMC) sampling approach is also considered, and the parameter estimates are based on the posterior mean.

For the multivariate normal distribution of Eq.~\eqref{eq:mvnorm}, if the data covariance $\mat C$ does not depend on $\vec \theta$, the
Fisher matrix $\mat F$ is given as \citep{1995PhDT........19B}
\begin{align}
	\mat F & =
	    \mat M^{\textrm{t}}
	   	\mat \Psi
	   	\mat M
        \label{eq:fish}
\end{align}
with $M_{ij} = \partial m_i / \partial \theta_j$,
see \citet{TTH97} for a seminal discussion in a cosmology context.
When the true precision matrix $\mat\Psi$ is replaced by
$\hat{\mat \Psi}_\ast$ and $\hat{\mat \Psi}$ (see Table~\ref{tab:estimators}), the corresponding estimated Fisher matrices are denoted as $\hat{\mat F}_\ast$ and $\hat{\mat F}$, respectively. The parameter covariance (estimate) is obtained by inverting the Fisher matrix (estimate), see App.~\ref{sec:phi} for more details.

The likelihood is no longer multivariate normal if the true precision matrix $\mat
\Psi$ is replaced by an estimate. 
Instead, a Hotelling $T^2$ likelihood function is appropriate \citep[see][]{hotelling1931,Mardia79}.
This distribution was also derived in \citet{2016MNRAS.456L.132S}, where they
marginalise the multivariate normal likelihood over the Wishart distribution of
the estimated inverse covariance matrix. The resulting log-likelihood is
\begin{equation}
  - \ln L_{T^2}(\vec y | \vec \theta) = \, {\rm const} + \frac {\nu + 1} 2 \ln \left[ 1 + \frac{\chi^2_\ast(\vec y | \vec \theta)}{\nu} \right] ,
  \label{eq:logT2}
\end{equation}
where $\chi^2_\ast$ is similar to the term given in Eq.~\eqref{eq:mvnorm}, but with
the estimated (biased) inverse covariance $\hat{\mat \Psi}_\ast$ instead of $\mat \Psi$.

For our first simulation study (Sect.~\ref{sec:toy_example}), we explore the multivariate normal of Eq.~\eqref{eq:mvnorm} and Hotelling $T^2$ likelihood of Eq.~\eqref{eq:logT2} with a Metropolis-Hastings Monte-Carlo Markov Chain
 sampler, implemented in \texttt{stan} \citep{JSSv076i01}. For each
number of simulations $n_{\rm s}$ used for the covariance estimation, we
produce $n_{\rm r}=50$ independent MCMC runs. We compute the convergence of the
samples by running three chains with $2,000$ points for each run, after
discarding the first $1,000$ burn-in phase chain points.

\subsection{Experiment: affine function with diagonal covariance matrix}
\label{sec:toy_example}

With this experiment we explore the capability of the proposed ABC algorithm to infer parameters compared to likelihood-based approaches using a model with a diagonal covariance matrix.

\subsubsection{Data-generating model}
\label{sec:exp1_data_gen}

We consider an affine function model $\vec m(\vec \theta) = a \vec x + b \vec 1$,
where the slope $a$ and intercept $b$ are the model parameters $\vec \theta = (a, b)$.
The input model is set to $\theta_0 = (1, 0)$.
As described in Sect.~\ref{sec:set_up_data_gen}, we fix the abscissa vector $\vec x$ and covariance matrix $\mat C$.
The former are generated once by drawing $p = 750$
uniform variables $x_i, i=1 \ldots p$,
$x_i \sim {\cal U}(-\Delta/2; \Delta/2)$ with $\Delta = 200$.
Note that this precludes the use of the acf distance, which requires an equidistantly spaced $\vec x$.
The input covariance matrix is the diagonal, matrix 
\begin{equation}
  \mat C = \sigma^2 \mat I_p.
  \label{eq:C_toy}
\end{equation}
Here, $\mat I_p$ is the $p \times p$ identity matrix, and we fix the value of the variance as $\sigma^2 = 5$.

\subsubsection{Distance function}
\label{sec:exp1_distance}

We use the parameter summary statistic of Eq.~\eqref{eq:s_param} and the corresponding distance function, $D_{\text{param}}$ from Eq.~\eqref{eq:dist_param}, with parameter $\vec \theta = (a, b)$.
We carry out one  modification to the distance function since we found that using the absolute value of $b$ improves the convergence and the resulting errors on $b$.
The \textit{modified parameter distance} is then
\begin{align}
  D^\prime_{\rm param}&[\vec s_{\rm param}(\vec d\Obs), \vec s_{\rm param}(\vec d\Sim)]
  = \sqrt{(a\Obs^\ast - a\Sim^\ast)^2 + (|b\Obs^\ast| - |b\Sim^\ast|)^2}.
    \label{eq:dist_param_ab}
\end{align}

\begin{figure}

\vspace*{-2em}

\begin{center}
\resizebox{\textwidth}{!}{
\includegraphics{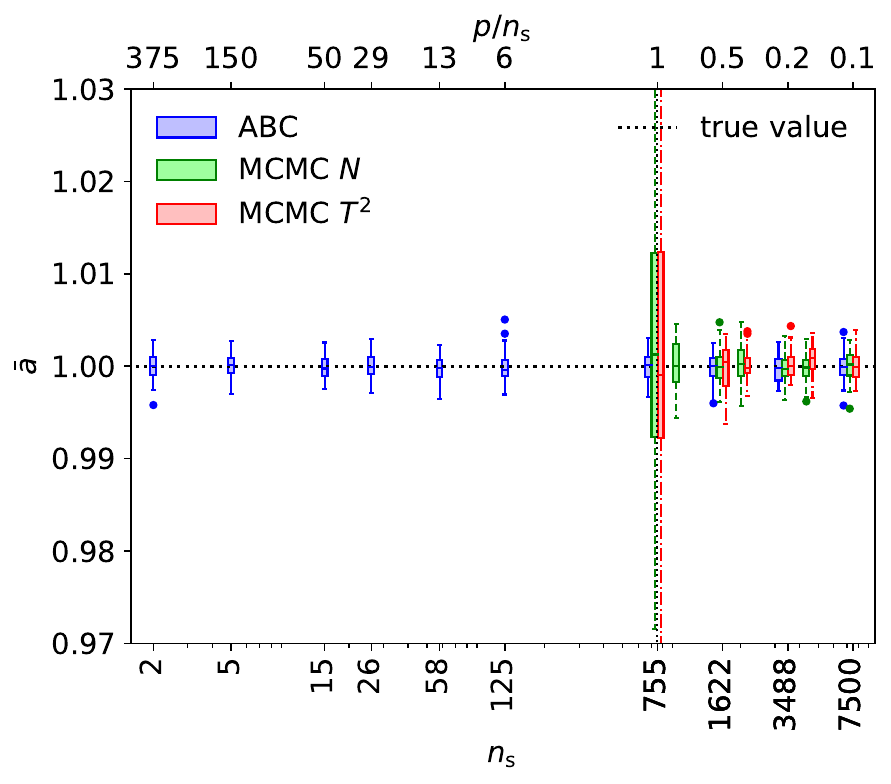}%
\includegraphics{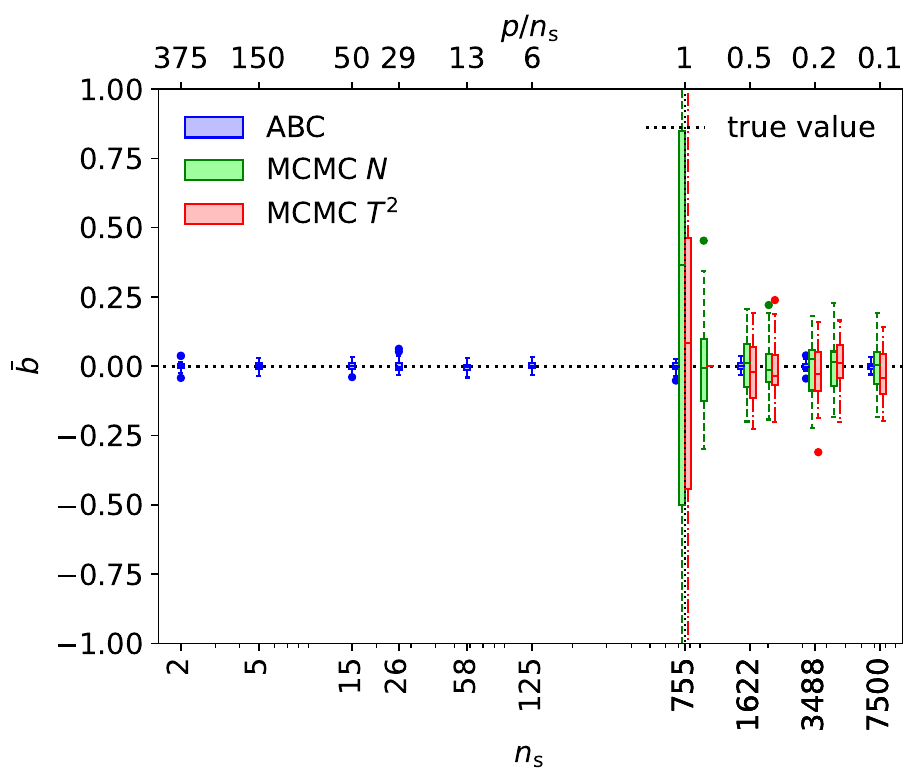}
}

\resizebox{\textwidth}{!}{
\includegraphics{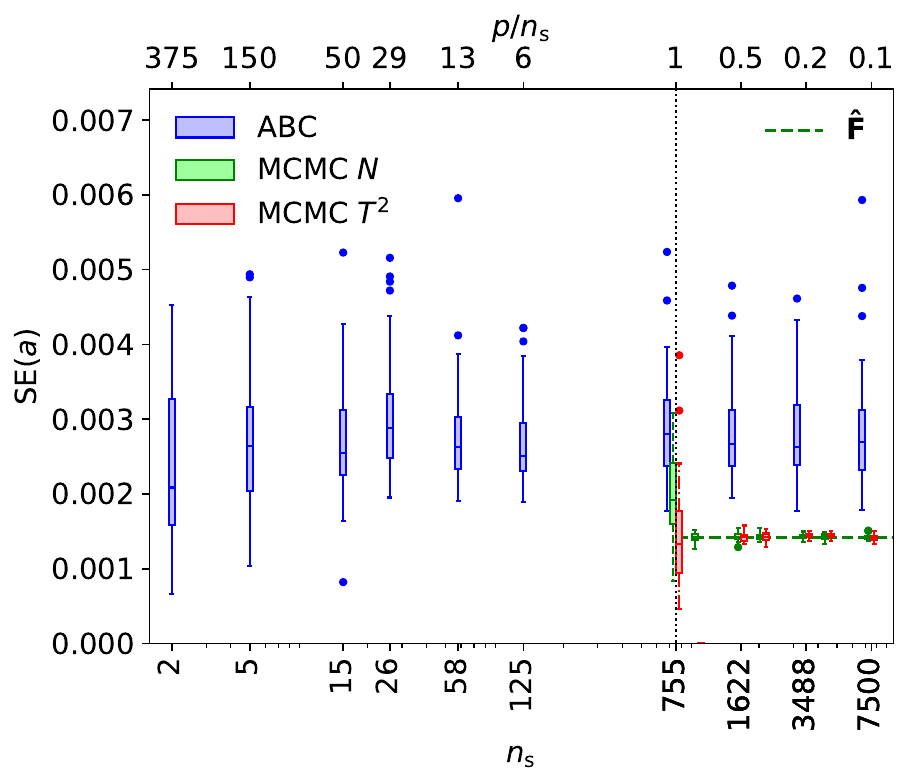}%
\includegraphics{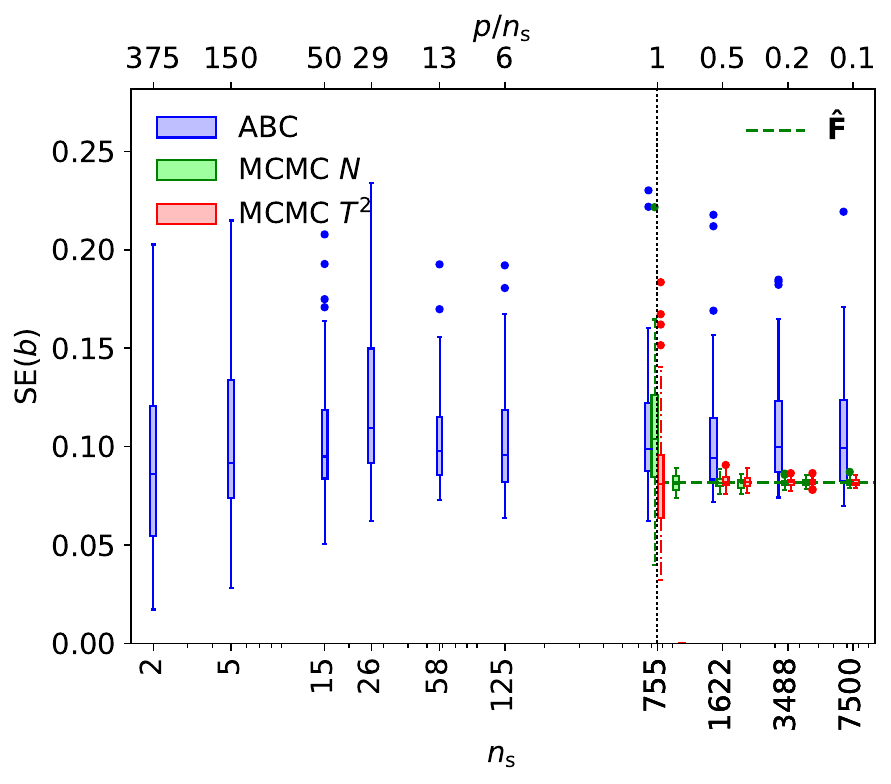}
}

\resizebox{\textwidth}{!}{
\includegraphics{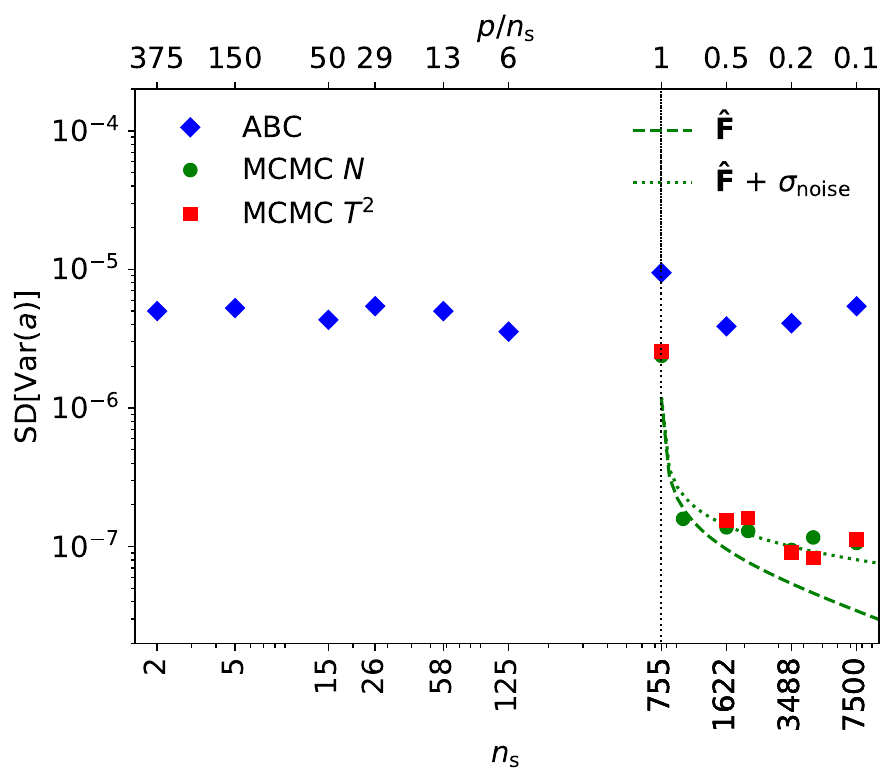}%
\includegraphics{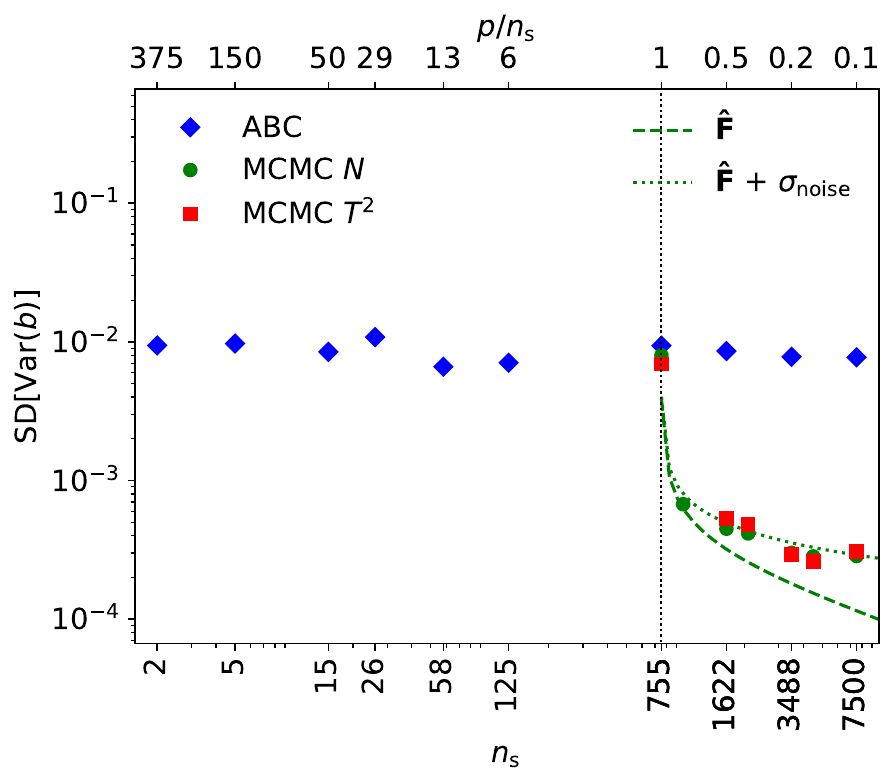}
}
\vspace*{-2em}
\end{center}

    \caption{Mean (\textit{top row}), SE (\textit{middle}) and SD of the
variance (\textit{bottom}) of the parameter estimates of $a$ (\textit{left
column}) and $b$ (\textit{right column}). Results are shown for ABC (solid blue
boxes, diamonds), and MCMC sampling of the normal ($T^2$) likelihood with
dashed green lines and circles (dash-dotted red lines and squares). Each point
corresponds to a number of simulations $n_{\textrm{s}}$ used for the covariance
estimate, and its inverse for MCMC sampling. The top panels show the true input
values for $a$ and $b$ as dotted lines. The middle panels display Fisher-matrix
prediction for the normal likelihood with dashed lines.
The bottom plots show as dashed line the prediction for the normal likelihood
(\ref{sec:sd_var_fisher}) and
the dotted line has added Monte-Carlo noise (see \S\ref{sec:exp1_results}).}

    \label{fig:results_te1}

\end{figure}

\subsubsection{Experiment details}
\label{sec:exp1_exp_details}

We use $M=100$ points in each
PMC iteration after the first, and a
convergence criterion of $\delta = 0.025$. This leads to a typical number of $\sim36$ iterations. The total number of proposed particles, for which the distance
needs to be computed, is around $30,000$, corresponding to a overall acceptance
rate\footnote{The number of accepted particles from all iterations divided by the total number of draws}
of $0.12$.
For comparison, we explore the multivariate normal likelihood from Eq.~\eqref{eq:mvnorm} and Hotelling $T^2$ likelihood from Eq.~\eqref{eq:logT2} with a Metropolis-Hastings MCMC sampler as discussed in Sect.~\ref{sec:mcmc}.

\subsubsection{Results}
\label{sec:exp1_results}

Fig.~\ref{fig:results_te1} shows the average estimate $\bar{\vec \hat{\theta}}$, the standard error SE($\vec \hat \theta)$, and the standard deviation of the variance SD$[$Var$(\vec \hat \theta)]$ of the parameter $\vec \theta = (a, b)$, for ABC and the likelihood-based comparison methods. From the
$n_{\textrm{r}} = 50$ runs we obtain the distribution of the mean and standard errors, shown as box plot. We estimate SD$[$Var$(\vec \hat \theta)]$ as the standard deviation
over the $50$ estimates of Var($\vec \hat \theta)$. 
The plot suggests unbiased mean estimates for both $a$ and $b$ for all methods considered,
even for a singular covariance when $p \ge n_{\rm s}$ for the proposed ABC method. This is true down to the extreme case of $n_{\rm s}=2$, corresponding to a
ratio $p/n_{\rm s} = 375$.
The SE estimates for ABC are larger compared to the multivariate normal prediction, see Sects.~\ref{sec:mcmc} and  \ref{sec:Fisher_matrix_ex_1}. No significant
dependence on the number of simulations $n_{\rm s}$ for the covariance
computation is visible.

For the normal and $T^2$ likelihood the estimated SD$[$Var$(\hat{\vec \theta})]$ is biased high
compared to the Fisher-matrix prediction, derived in \ref{sec:sd_var_fisher}. There is additional variance intrinsic to
Monte-Carlo sampling that does not stem from the inversion of the covariance.
We estimate this uncertainty by evaluating the normal likelihood function with
the true precision matrix $\mat \Psi = \sigma^{-2} \mat I_p$. We find a
distribution of the variances of the parameter estimators
with finite width, reflecting the sampling noise,
which we estimate 
as SD$_\mathrm{noise}[$Var$(\hat{a})] = 4.6 \times
10^{-8}$, and SD$_\mathrm{noise}[$Var$(\hat{b})] = 1.75 \times 10^{-4}$, with around $10\%$
uncertainty on these values. When this is added to the predicted
SD[Var] values (dotted lines in
Fig.~\ref{fig:results_te1}), there is good agreement with the Fisher-matrix
prediction. 

The normal likelihood has a smaller estimated SD$[$Var$(\hat{\vec \theta})]$ compared to ABC, and decreases with increasing $n_\mathrm{s}$. 
The SE of the estimators diverge for both the normal and the Hotelling $T^2$ likelihood at $n_{\rm s} = p$.
The SD$[$Var$(\hat{\vec \theta})]$ from ABC are larger than the ones corresponding to sampling under  either likelihood, but do not show a significant dependence on $n_\mathrm{s}$ remaining more or less constant down to $n_\mathrm{s} = 2$. 

\subsection{Experiment: a weak-gravitational-lensing inspired case}
\label{sec:WL_example}

The model for the experiment discussed in this section approximates more closely the statistical
properties of weak-gravitational lensing data, in particular the cosmic shear
power spectrum observable, $C_\ell$, where $\ell$ is the 2D Fourier wave number
on the sky.  In this study, two distance function options are considered to compare their performances, including assessing the usefulness of accounting for correlations in the data points.

\subsubsection{Data-generating model}
\label{sec:exp2_data_gen}

\begin{SCfigure}
    \includegraphics[scale=.65]{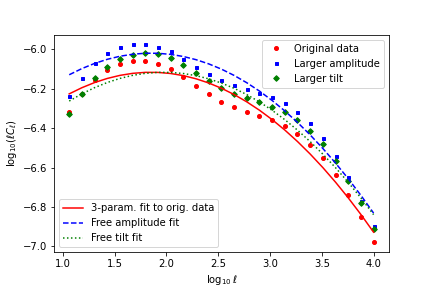}
     \caption{Weak-lensing power spectrum $C_\ell$ (points),
     and best-fit quadratic functions (lines) from Eq.~\eqref{eq:qu}. The red circles correspond to
     parameters $\Omega_{\rm m} = 0.306$ and
     $\sigma_8 = 0.827$. For the blue squares (green diamonds), the values
     of $\sigma_8$ ($\Omega_{\rm m}$) are increased by $10\%$. The dashed and
     dotted lines are the best-fit curves with free amplitude and tilt,
     respectively.
     \nblink{ABC/WL_Cell_fit}
     }

     \label{fig:qu_pm_fit}
\end{SCfigure}

This example is an analytical model that mimics a weak-lensing power spectrum $C_\ell$
(Sect.~\ref{sec:WL_example}). With respect to the previous example (Sect.~\ref{sec:toy_example}), additional complexity  typically arising for weak lensing is included: First, the model is non-linear in the parameters, and second, data points are correlated. 
We follow Sect.~\ref{sec:set_up_data_gen} to generate the data vectors for this experiment. The following quadratic function is considered
\begin{equation}
  q(x) = c + a (x - x_0)^2,
  \label{eq:qu}
\end{equation}
which is a rough approximation of $\LG \left( \ell C_\ell \right)$, where we identify $x \equiv \LG \ell$. In Fig.~\ref{fig:qu_pm_fit},
we fit the parabola of Eq.~\eqref{eq:qu} to a weak-lensing
power spectrum model, obtained
with the software \texttt{nicaea}\footnote{https://github.com/CosmoStat/nicaea} \citep{KB09}.

Two parameters are defined that reproduce the effect of the main cosmological
parameters on $C_\ell$, as follows.
First, a tilt parameter $t$ corresponds to the matter density $\Omega_{\rm m}$.
The matter density determines the epoch of matter-radiation equality in the
early Universe, which is responsible (among other factors) for the peak in $C_\ell$.
For a larger $\Omega_{\rm m}$, the matter-dominated phase starts earlier, and
suppression of power in the radiation-dominated era on small scales (large
$\ell$) is reduced, shifting the peak to the right. The tilt is proportional to
the shift parameter $x_0$ in Eq.~\eqref{eq:qu}.
Second, an amplitude parameter $A$ mimics the 3D density power-spectrum
normalisation $\sigma_8$. To first-order, $C_\ell \propto \sigma_8^2$, and thus
$A$ is given by twice the logarithm of the constant term $c$ in Eq.~\eqref{eq:qu}.

Choosing the proportionality constants between $t, A$, and the parabola
parameters such that the best-fit values of $t$ and $A$ corresponds to the
input model parameters $\Omega_{\rm m} = 0.306$ and  $\sigma_8 = 0.827$, we find
\begin{equation}
  x_0 = 6.05 \, t; \quad\quad c = 2 \LG A -5.95,
  \label{eq:u0_c}
\end{equation}
where $a = -0.176$ in Eq.~\eqref{eq:qu} is fixed. Fig.~\ref{fig:qu_pm_fit} shows that changing
$\Omega_{\rm m}$ and $\sigma_8$ by a small amount can be reproduced by changing
$t$ and $A$ by similar amounts, without modifying the other parameters.

Such an approximation would not be acceptable for cosmological modeling,
but is sufficient for this experiment because we are mainly interested in a simple and
analytical example that roughly reproduces cosmological effects.

The abscissa $\vec x$ of the data vector (see Sect.~\ref{sec:s_d}) consists of $p = 10$ values $x_i
= \ell_i$, equally spaced in $\ell$ between $\ell_{\rm min} = 25$ and
$\ell_{\rm max} = 4,000$. The ordinate vector is chosen to mimic the $C_{\ell}$. In this approximation for the weak-lensing power spectrum, this corresponds to
\begin{equation}
  m_i = 10^{q(\LG \ell_i)} \ell_i^{-1} = 10^{q(x_i) - x_i}.
  \label{eq:y_tm3}
\end{equation}
As described in Sect.~\ref{sec:set_up_data_gen}, $n_{\textrm{r}}$ independent observable vectors are generated as multivariate normal, $y\Obs^r \sim {\cal N}(\vec m(\vec \theta_0), \mat C)$.
The fixed model parameter vector is $\vec \theta_0 = (t_0 = 0.306, A_0 = 0.827)$.

Two scenarios are used for the uncertainty and correlations between data points
in our model. First, we use the diagonal Gaussian covariance matrix
with elements according to Eq.~\eqref{eq:cov_WL_G}, where the data points
are uncorrelated. Then we consider the Gaussian plus SSC covariance 
(see Sect.~\ref{sec:WL_cov})
to model the non-Gaussian and non-linear evolution of the weak-lensing power spectrum on small scales.

For the Gaussian term we mimic the weak-lensing Gaussian covariance given by
Eq.~\eqref{eq:C_obs} and Eq.~\eqref{eq:cov_WL_G}, where we insert $y$ from Eq.~\eqref{eq:y_tm3}
for the ``signal'' $C_\ell$.  For the SSC term, we use the one derived in
\citet{2018JCAP...06..015B}, and parameterize the SSC
contribution scaled by the weak-lensing power spectrum, see Fig.~3 in
\citet{2018JCAP...06..015B}. The correlation matrix of the total covariance
(Gaussian + SSC) is plotted in Fig.~\ref{fig:r_G+SSC}. 

To set the numerical values in the covariance matrix, Eqs.~\eqref{eq:C_obs} and \eqref{eq:cov_WL_G}, we model a survey with properties similar to what is expected for Euclid \citep{2011arXiv1110.3193L,Euclid-prep-VII_ISTF_20} with
$f_{\textrm{sky}} = 0.3636$, corresponding to a observed sky area of $15,000$
deg$^2$. The galaxy density in the single redshift bin is chosen as $\bar n =
30$ arcmin$^{-2}$, the intrinsic ellipticity dispersion set to
$\sigma_\varepsilon = 0.31$.

\begin{SCfigure}
 		\includegraphics[viewport=65 0 310 216]{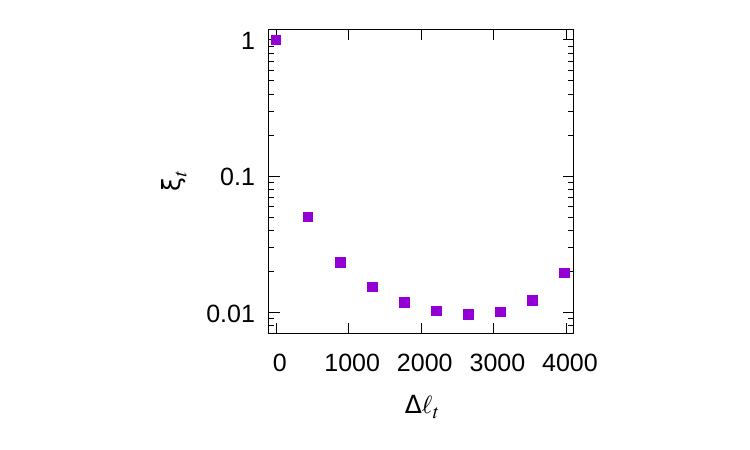}
     \caption{A realization of the normalized auto-correlation function $\xi_t$ (\ref{eq:xi}).
     The x-axis shows the corresponding angular scale difference $\Delta \ell_t = \ell_{i+t} - \ell_{i}$,
     which is independent of $i$.
     }
     \label{fig:xi}
\end{SCfigure}

\subsubsection{Distance function}
\label{sec:quad_distance}

\begin{SCfigure}
 		\includegraphics[scale=0.6]{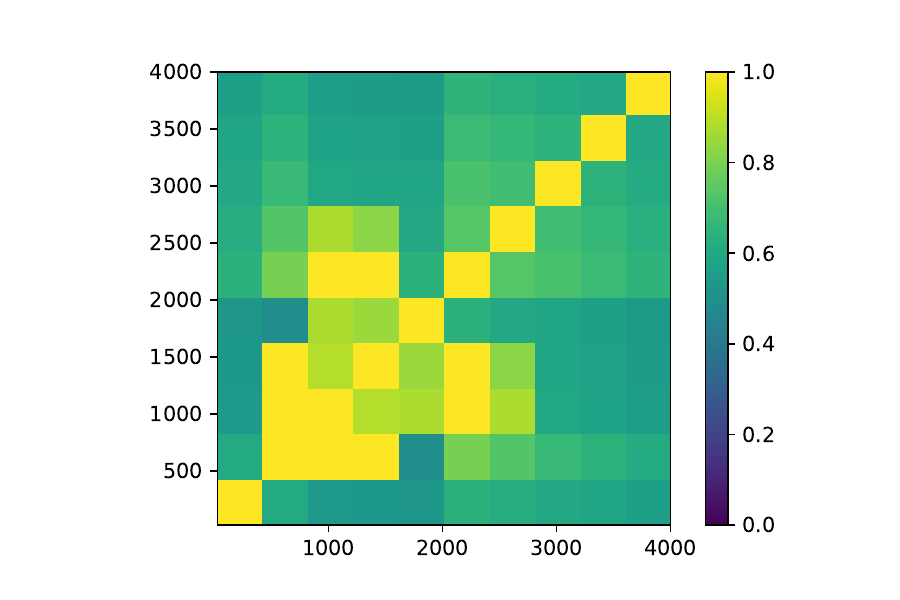}
     \caption{Correlation coefficient of the total covariance matrix (Gaussian + SSC).
     The axis labels show the values of $\ell$.
     }
     \label{fig:r_G+SSC}
\end{SCfigure}

Because the data are correlated in this setting, we consider the 
acf distance function of Eq.~\eqref{eq:d_acf}. The results are compared with the inverse-variance distance of Eq.~\eqref{eq:dist_data_diag}, which ignores the correlation in the data.
The acf of one of the $n_\textrm{r}=25$ realizations of the  observations $\vec y\Sim$ is displayed in Fig.~\ref{fig:xi}. After a sharp drop from zero lag, the
correlation stays above zero. This represents the correlation between small and large scales (large $\Delta \ell_t$), which can be seen in
the covariance matrix (Fig.~\ref{fig:r_G+SSC}). The acf can thus capture some of the correlation information of the data vector, to be used in the distance function.

\subsubsection{Experiment details}

As in the previous example, each PMC iteration after the first is carried out with $M = 100$ accepted particles. To reach convergence, we use a slightly tighter convergence criterion of $\delta = 0.015$ resulting in a mean number of iterations to reach convergence of $32$. The total number of draws per run is $40,000$, corresponding to an overall acceptance rate of around $0.08$.
The focus of this experiment is the comparison of different ABC distances regarding correlations in the data.
Sect.~\ref{sec:exp1_results} has established the accuracy of our Fisher-matrix prediction by comparing those with MCMC sampling, which we do not repeat here. Instead, we only
compare ABC to the Fisher matrix, see \ref{sec:app_phi_WL_example} for the expressions for this example.

\begin{table}[bh!]
  \centering
 
\begin{tabular}{|c|c|c|c|ccc|ccc|}
  \hline
$\mat C$ &
dist. & infer. & $n_{\textrm{s}}$ &
$\bar{\hat t}$ & $\!\!$SE$(\hat t)$ & $\!\!\!\!$SD$[$Var$(\hat t)]$ & $\bar{\hat A}$ & $\!\!$SE$(\hat A)$ & $\!\!\!\!$SD$[$Var$(\hat A)]$ \\
\hline\hline
\multirow{2}{*}{G} & \multirow{2}{*}{$\hat{\mat \Psi}$} & \multirow{2}{*}{$\hat{\mat F}$} & $\le p$ &
    - & \multicolumn{2}{c|}{not computable} &
    - & \multicolumn{2}{c|}{not computable} \\
& & & $>p$ &
    - & $0.0007$ & $2.9 \cdot 10^{-7}$ &
    - & $0.0019$ & $2.2 \cdot 10^{-6}$ \\ \hline
%\multirow{2}{*}{G} & \multirow{2}{*}{$\hat{\mat \Psi}_\ast$} & %\multirow{2}{*}{$\hat{\mat F}_{T^2}$} & $\le p$ &
%    - & \multicolumn{2}{c|}{not computable} &
%    - & \multicolumn{2}{c|}{not computable} \\
%& & & $>p$ &
%    - & $0.0010$ & - &
%    - & $0.0027$ & - \\ \hline
%G     & diag $\!\hat{\mat C}$   & ABC & all     & $0.3059$ & $0.0007$ & $2.4 \cdot 10^{-7}$ & $0.8275$ & $0.0020$ & $1.8 \cdot 10^{-6}$ \\
\multirow{2}{*}{G} & \multirow{2}{*}{diag $\!\hat{\mat C}$} & \multirow{2}{*}{ABC} & $\le p$ &
    \hi $0.3058$ & \hi $0.0007$ & \hi $2.4 \cdot 10^{-7}$ &
    \hi $0.8275$ & \hi $0.0019$ & \hi $1.8 \cdot 10^{-6}$ \\
& & & $> p$ &
    $0.3059$ & $0.0008$ & $2.3 \cdot 10^{-7}$ &
    $0.8275$ & $0.0022$ & $2.0 \cdot 10^{-6}$ \\ \hline\hline
\multirow{2}{*}{G+SSC} & \multirow{2}{*}{$\hat{\mat \Psi}$} & \multirow{2}{*}{$\hat{\mat F}$} & $\le p$ &
    - & \multicolumn{2}{c|}{not computable} &
    - & \multicolumn{2}{c|}{not computable} \\
& & & $> p$ &
    - & $0.0007$ & $3.3 \cdot 10^{-7}$ &
    - & $0.0039$ & $9.1 \cdot 10^{-6}$ \\ \hline
%\multirow{2}{*}{G+SSC} & \multirow{2}{*}{$\hat{\mat \Psi}_\ast$} & %\multirow{2}{*}{$\hat{\mat F}_{T^2}$} & $\le p$ &
%    - & \multicolumn{2}{c|}{not computable} &
%        - & \multicolumn{2}{c|}{not computable} \\
%& & & $> p$ &
%    - & $0.0011$ & - &
%    - & $0.0056$ & - \\ \hline
%
%G+SSC &  diag $\!\hat{\mat C}$  & ABC & all     & $0.3060$ & $0.0012$ & $6.0 \cdot 10^{-7}$ & $0.8263$ & $0.0046$ & $9.4 \cdot 10^{-6}$ \\
\multirow{2}{*}{G+SSC} & \multirow{2}{*}{diag $\!\hat{\mat C}$} & \multirow{2}{*}{ABC} & $\le p$ &
    \hi $0.3060$ & \hi $0.0012$ & \hi $5.3 \cdot 10^{-7}$ &
    \hi $0.8265$ & \hi $0.0043$ & \hi $9.8 \cdot 10^{-6}$ \\
& & & $> p$ &
    $0.3061$ & $0.0013$ & $7.2 \cdot 10^{-7}$ &
    $0.8260$ & $0.0050$ & $8.7 \cdot 10^{-6}$ \\ 
\hline
%
%G+SSC & acf                     & ABC & all     & $0.3056$ & $0.0024$ & $2.1 \cdot 10^{-6}$ & $0.8266$ & $0.0051$ & $1.2 \cdot 10^{-5}$ \\
\multirow{2}{*}{G+SSC} & \multirow{2}{*}{acf} & \multirow{2}{*}{ABC} & $\le p$ &
    \hi $0.3056$ & \hi $0.0024$ & \hi $2.4 \cdot 10^{-6}$ &
    \hi $0.8266$ & \hi $0.0049$ & \hi $1.3 \cdot 10^{-5}$ \\
& & & $> p$  &
    $0.3055$ & $0.0025$ & $1.6 \cdot 10^{-6}$ &
    $0.8266$ & $0.0054$ & $1.1 \cdot 10^{-5}$ \\ 
  \hline
\end{tabular}
\caption{Results for the weak-lensing inspired example of Sect.~\ref{sec:WL_example}. 
\textit{Columns:}
$\mat C$: The covariance matrix used in the model, Gaussian (G), or Gaussian plus super-survey covariance (G+SSC);
\textit{dist}: The covariance component of the distance function; the corresponding equations are Eq.~\eqref{eq:fish} for the Fisher matrix of the normal likelihood; Eq.~\eqref{eq:dist_data_diag} for the inverse-variance distance; and Eq.~\eqref{eq:d_acf} for the acf distance, which does not depend on the covariance matrix;
\textit{infer.}: Inference method;
$n_\textrm{s}$: Simulation size compared to the data vector length $p$.
$\hat t$, $\hat A$: Mean of tilt and amplitude estimates.
SE$(\hat t)$, SE$(\hat A)$: Standard error of tilt and amplitude estimate.
SD$[$Var$(\hat t)]$,  SD$[$Var$(\hat A)]$: Standard deviation of the variance of 
tilt and amplitude estimates;
The true input values are $t=0.306$ and $A=0.827$. For the Fisher matrix we use Eq.~\eqref{eq:SDVarF} to obtain SD[Var].
Highlighted cells correspond to a singular covariance.
}
\label{tab:results_exp2}
\end{table}

\subsubsection{Results}

\begin{figure}
  \begin{center}
 		\resizebox{1.0\columnwidth}{!}{
 		  \includegraphics{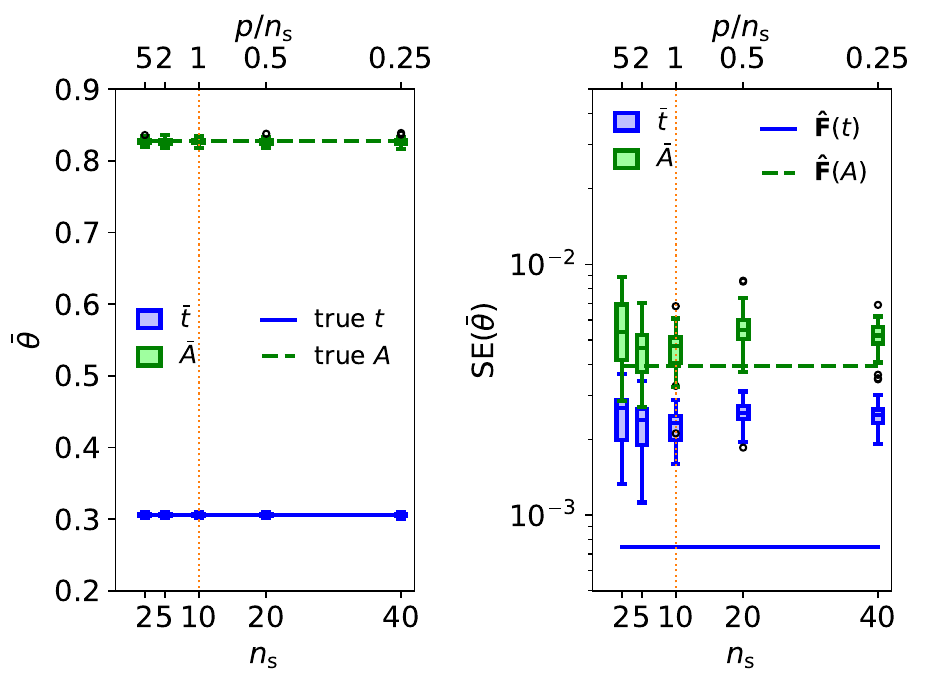}
 		  \raisebox{-1.0ex}{\includegraphics{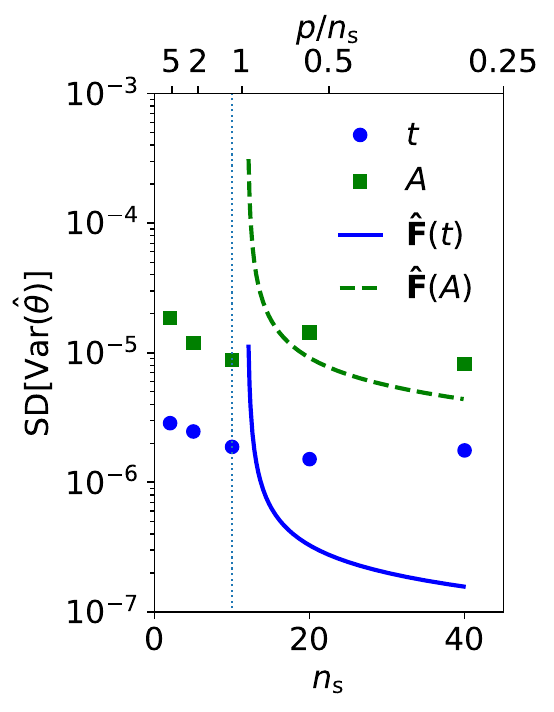}}
     }
  \end{center}

     \caption{Mean (\textit{left}), standard error (\textit{middle}), and standard deviation of the variance (\textit{right}) of the
     parameters estimates for $t$ and $A$ from the weak-gravitational-lensing inspired model of Sect.~\ref{sec:WL_example}. The acf
     distance of Eq.~\eqref{eq:d_acf} is used. Each simulation size $n_\textrm{s}$ corresponds to a number of
     simulations $n_{\rm s}$ used for the covariance matrix estimate.
     In the left plot, horizontal lines show the true input values.
     The middle and right panels display predictions based on the Fisher matrix
     for the normal likelihood (thin lines).
     Average values are also displayed in Table \ref{tab:results_exp2}.
     }

     \label{fig:mean_std_ABC_quad}
\end{figure}

Fig.~\ref{fig:mean_std_ABC_quad} shows ABC under the non-Gaussian covariance model (G+SSC) with the acf distance Eq.~\eqref{eq:d_acf}.
In all cases the ABC parameter estimates are consistent with the true values.  
The results are summarized in Table \ref{tab:results_exp2} for different distance functions and compared using the Fisher-matrix estimates. We average over $n_{\textrm{r}} = 25$ runs, and two ranges of the number of simulations: (1) the values $n_{\textrm{s}} = 2, 5, 10$, written as ``$\le p$'' in the table; and (2) the values $n_\textrm{s} = 20, 40$, denoted by ``$> p$''. For (2) the covariance matrix is non-singular, and thus only this case is available for the Fisher-matrix predictions.

For a Gaussian input model covariance
ABC (with the inverse-variance distance (Eq.~\eqref{eq:dist_data_diag}) shows standard errors (SE)
of the parameter estimates close to the Cram\'er-Rao bound.

With correlated data points via the addition of the SSC term,
under the normal likelihood, the SE for the estimated tilt $\hat t$ is similar than for the
uncorrelated case. However, the SE of the amplitude estimate $\hat A$ doubles.
With ABC and the inverse-variance distance the SE of $\hat t$
$\hat A$. Using the acf distance the SE($\hat t$) doubles,
but the SE$(\hat A)$ increases only by a small amount.

Table \ref{tab:results_exp2} also shows the standard deviation of the variance with the different $n_\textrm{s}$ cases.
The acf distance results in larger variations compared to the inverse-variance distance. 
No systematic increase is visible for the case of singular covariance,
and the values are relatively independent of $n_{\textrm{s}}$, see Fig.~\ref{fig:mean_std_ABC_quad}. This is not the case for the Fisher-matrix prediction, as already seen in the previous example, and SD[Var] diverges for $n_\textrm{s} = p + 2$.

To summarize, the acf distance function results in a SE and a SD of the variance of the parameter estimate that are larger compared to the inverse-variance distance. For both distance, the errors of the parameter estimates are above the Cram\'er-Rao bound.

\subsection{Experiment: A realistic weak-gravitational lensing model}
\label{sec:WL_model}

In this section a realistic weak-lensing power spectrum $C_\ell$ derived from a numerical, non-linear model of the large-scale structure and lensing projection is considered \citep{2012ApJ...761..152T}. The free parameters are the matter density $\Omega_\textrm{m}$ and the power-spectrum normalisation $\sigma_8$.
One of the main astrophysical contaminants to weak lensing is intrinsic galaxy alignment. Galaxy shapes can be correlated to their surrounding dark-matter
environment by gravitational interactions. Alignments can be created by the exertion of torquing moments, or anisotropic stretching and accretion,
induced by the surrounding tidal field. Intrinsic alignment creates correlations between galaxy intrinsic galaxy ellipticities (II),
and between shear and intrinsic galaxy ellipticities (GI).
The former is only important for galaxies very close in redshift, the relative number of which is low in our case of a single broad redshift distribution.
Therefore we only account for the GI cross-correlation.
Intrinsic alignment can be modeled as a power spectrum $C_\ell^{\textrm{GI}}$, which is added to the weak-lensing power spectrum defined in Eq.~\eqref{eq:p_kappa_limber} with an amplitude $A_\textrm{IA}$. This results in a new expression for the observed power spectrum from Eq.~\eqref{eq:C_obs} as
\begin{equation}
C_{\ell, {\rm obs}} = C_\ell + A_\textrm{IA} C_\ell^{\textrm{GI}}
+ \frac{\sigma_\varepsilon^2}{\bar n}.
\end{equation}
In this experiment, we select a fixed value of $A_\textrm{IA} = 2$, which follows \cite{2021MNRAS.501.2983F}.

\subsubsection{Data-generating model}
\label{sec:wl_data_gen}

As in the previous example the data is composed of the abscissa vector $\vec x$ consisting of  $ n_\mathrm{d}= 10$ values $x_i
= \ell_i$, equally spaced in $\ell$ between $\ell_{\rm min} = 25$ and
$\ell_{\rm max} = 4,000$. The ordinate vector is the weak-lensing power spectrum,  $\vec y = \{C_{\ell_i}\}$.
The covariance matrix is the same as in Sect.~\ref{sec:exp2_data_gen}; we use the two cases of a Gaussian, and of a Gaussian plus SSC covariance.
We chose again a set-up corresponding to a survey similar to Euclid, with observed sky area fraction $f_\textrm{sky} = 0.363$,  galaxy density $\bar n = 30$ arcmin$^{-2}$, and intrinsic ellipticity dispersion $\sigma_\varepsilon = 0.31$.

\subsubsection{Distance function.} We consider the  inverse-variance distance Eq.~\eqref{eq:dist_data_diag}, and the acf distance Eq.~\eqref{eq:d_acf}.

\begin{table}[ht!]

\begin{tabular}{|c|c|c|c|ccc|ccc|}
\hline
$\mat C$ & dist. & infer. & $n_{\textrm{s}}$  &
$\overline{\hat \Omega}_\mathrm{m}$ & $\!\!$SE$(\hat \Omega_\mathrm{m})$ & $\!\!\!$SD$[$Var$(\hat \Omega_\mathrm{m})]$ & $\overline{\hat \sigma}_8$ & $\!\!$SE$(\hat \sigma_8)$ & $\!\!\!$SD$[$Var$(\hat \sigma_8)]$ \\ \hline
\multirow{2}{*}{G} & \multirow{2}{*}{$\hat{\mat \Psi}$} & \multirow{2}{*}{$\hat{\mat F}$} & $\le p$ &
    - & \multicolumn{2}{c|}{not computable} &
    - & \multicolumn{2}{c|}{not computable} \\
& & & $> p$ &
    - & $0.0032$ & $6.0 \cdot 10^{-6}$ &
    - & $0.0047$ & $1.3 \cdot 10^{-5}$ \\ \hline
%\multirow{2}{*}{G} & \multirow{2}{*}{$\hat{\mat \Psi}_\ast$} & \multirow{2}{*}{$\hat{\mat F}_{T^2}$} & $\le p$ &
    %- & \multicolumn{2}{c|}{not computable} &
    %- &  \multicolumn{2}{c|}{not computable} \\
%&  & & $> p$ &
    %- & $0.0045$ & - &
    %%- & $0.0066$ & - \\ \hline
%G     & diag $\!\hat{\mat C}$ & ABC & all & $0.3063$ & $0.0034$ & $2.1 \cdot 10^{-6}$ & $0.8273$ & $0.0050$ & $5.5 \cdot 10^{-6}$ \\ 
\multirow{2}{*}{G}     & \multirow{2}{*}{diag $\!\hat{\mat C}$} & \multirow{2}{*}{ABC} & $\le p$ &
    \hi $0.3060$ & \hi $0.0038$ & \hi $6.7 \cdot 10^{-6}$ &
    \hi $0.8268$ & \hi $0.0056$ & \hi $1.4 \cdot 10^{-5}$ \\
& & & $> p$ &
    $0.3057$ & $0.0038$ & $4.8 \cdot 10^{-6}$ &
    $0.8273$ & $0.0056$ & $1.0 \cdot 10^{-5}$ \\ \hline\hline
\multirow{2}{*}{G+SSC} & \multirow{2}{*}{$\hat{\mat \Psi}$} & \multirow{2}{*}{$\hat{\mat F}$} & $\le p$ &
    - & \multicolumn{2}{c|}{not computable} &
    - & \multicolumn{2}{c|}{not computable} \\
& & & $> p$ &
    - & $0.0033$ & $6.2 \cdot 10^{-6}$ &
    - & $0.0060$ & $2.1 \cdot 10^{-5}$ \\ \hline
%\multirow{2}{*}{G+SSC} & \multirow{2}{*}{$\hat{\mat \Psi}_\ast$} & \multirow{2}{*}{$\hat{\mat F}_{T^2}$} & $\le p$ &
    %- & \multicolumn{2}{c|}{not computable} &
    %- & \multicolumn{2}{c|}{not computable} \\ 
 %& & & $> p$ &
    %- & $0.0046$ & - & 
    %- & $0.0084$ & - \\ \hline
%
%G+SSC & diag $\!\hat{\mat C}$ & ABC & all     & $0.3062$ & $0.0036$ & $5.3 \cdot 10^{-6}$ & $0.8264$ & $0.0061$ & $1.5 \cdot 10^{-5}$ \\
\multirow{2}{*}{G+SSC} & \multirow{2}{*}{diag $\!\hat{\mat C}$} & \multirow{2}{*}{ABC} & $\le p$ &
    \hi $0.3058$ & \hi $0.0036$ & \hi $5.6 \cdot 10^{-6}$ &
    \hi $0.8270$ & \hi $0.0061$ & \hi $1.6 \cdot 10^{-5}$ \\
 & & & $> p$   &
    $0.3067$ & $0.0036$ & $4.8 \cdot 10^{-6}$ &
    $0.8256$ & $0.0060$ & $1.3 \cdot 10^{-5}$ \\ \hline
%
%G+SSC & acf                   & ABC & all     & $0.3063$ & $0.0054$ & $1.2 \cdot 10^{-5}$ & $0.8265$ & $0.0076$ & $2.6 \cdot 10^{-5}$ \\
\multirow{2}{*}{G+SSC} & \multirow{2}{*}{acf} & \multirow{2}{*}{ABC} & $\le p$ &
    \hi $0.3064$ & \hi $0.0054$ & \hi $1.1 \cdot 10^{-5}$ &
    \hi $0.8264$ & \hi $0.0077$ & \hi $2.6 \cdot 10^{-5}$ \\
&  &  & $> p$   &
    $0.3062$ & $0.0055$ & $1.4 \cdot 10^{-5}$ &
    $0.8267$ & $0.0076$ & $2.5 \cdot 10^{-5}$ \\ \hline
\end{tabular}

\caption{Results for the realistic weak-lensing case of Sect.\ref{sec:WL_model}. See Table \ref{tab:results_exp2} for details. The true input parameters are $\Omega_\textrm{m} = 0.306$ and $\sigma_8 = 0.827$.}

\label{tab:results_wl}

\end{table}

\subsubsection{Experiment details}
\label{sec:wl_exp_details}

Each iteration is run with twice the number of points compared to the previous two examples, $M = 200$ points, and the initial number of draws from the prior is $M = 400$. The convergence criterion is $\delta = 0.015$, leading to $41$ iterations on average. With on average $87,000$ draws per run, the overall acceptance rate is $0.1$.

\subsubsection{Results}
\label{sec:wl_results}

\begin{figure}
  \begin{center}
 		\resizebox{1.0\columnwidth}{!}{
 		  \includegraphics{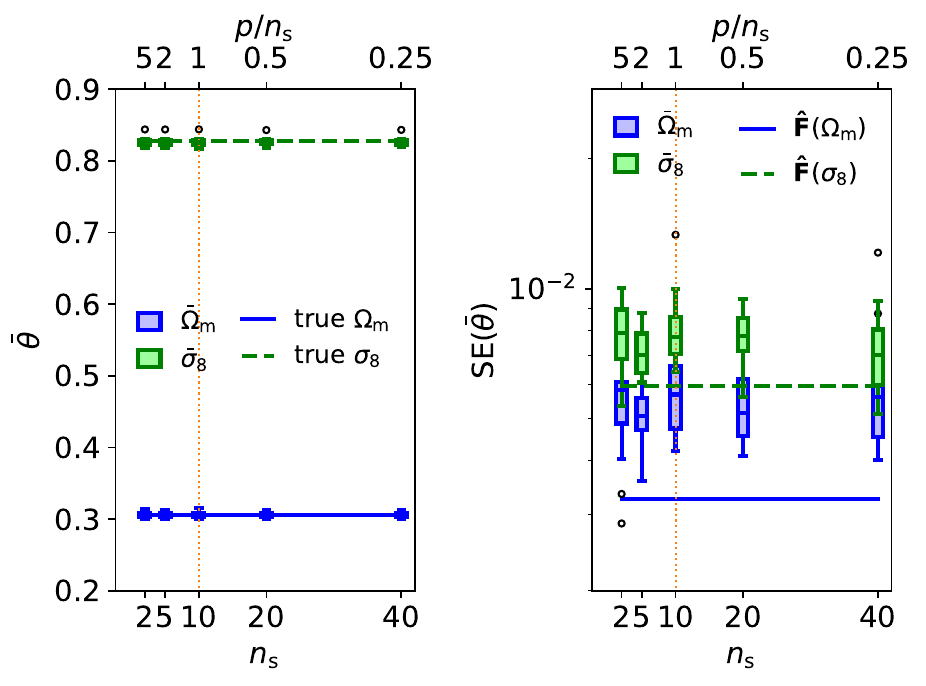}
	  \raisebox{-1.0ex}{\includegraphics{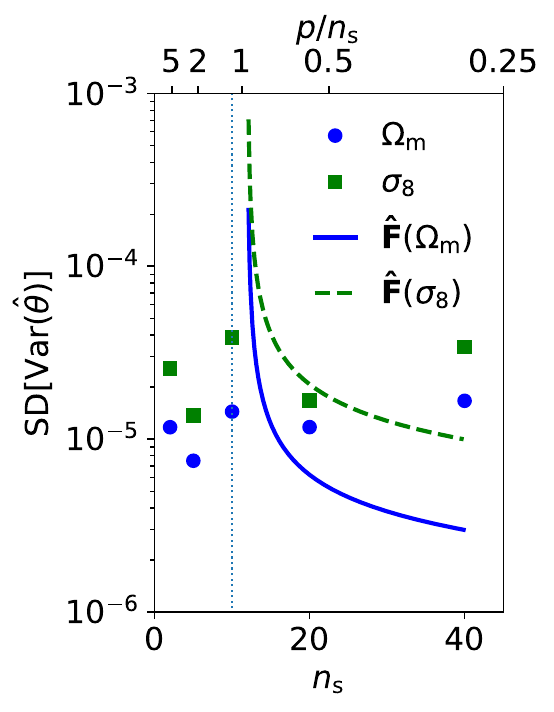}}
     }
  \end{center}
     \caption{Mean (\textit{left}), SE (\textit{middle}), and standard deviation of the variance (SD[Var]; \textit{right}) of the parameter estimates of $\Omega_\textrm{m}$ and $\sigma_8$ from the realistic weak-gravitational lensing model of Sect.~\ref{sec:WL_model}. The
     acf distance of Eq.~\eqref{eq:d_acf} is used.
     See Fig.~\ref{fig:mean_std_ABC_quad} for more details. 
     Numerical values
     for this setting are shown in Table \ref{tab:results_wl}.}
     \label{fig:mean_std_ABC_wl}
\end{figure}

The results are summarized in Table \ref{tab:results_wl}, and Fig.~\ref{fig:mean_std_ABC_wl} shows ABC with the acf distance compared to Fisher-matrix predictions.
The results for this weak-lensing case are similar to the quadratic model from Sect.~\ref{sec:WL_example}: the ABC estimates are consistent with the true parameter values. There is no visible dependence on the number of simulations used to compute the covariance matrix, including the case of a singular matrix.
We see no adversarial effects on the results from forward-modelling simulations generated with a small number of simulations and under a singular covariance.
The main difference to Sect.~\ref{sec:WL_example} is that SE and SD[Var] for $\hat \Omega_\textrm{m}$ from ABC is close to the normal likelihood case. The tilt parameter $t$ in Sect.~\ref{sec:WL_example} had values of SE and SD[Var] significantly larger than the Fisher-matrix predictions, contrary to ${\Omega}_\textrm{m}$ in this example.

For the inverse-variance distance the SE of the parameter estimates are very close to the Cram\'er-Rao bound. The SD[Var] are smaller than in the multivariate normal case. The results from the acf distance are never below the Cram\'er-Rao case.
This indicates that the inverse-variance distance under-estimates the parameter errors due to neglecting the correlation of the observed data vector. This is remedied with the acf distance.

\section{Conclusions}
\label{sec:conclusions}

This paper addresses the challenge of inference in the presence of a
singular covariance matrix estimate $\hat{\mat C}$. This can be the case for correlated cosmological
observations of the large-scale structure, where the covariance matrix is
estimated from $N$-body simulations of the data. If the
(typically small since computationally costly) number of simulations $n_{\textrm{s}}$ is less than the
(typically large) data dimension $p$, $\hat{\mat C}$ is singular.  Likelihood-based inference that requires a precision matrix estimate $\hat{\mat
\Psi}$ is not possible in that case. The situation is exacerbated if the
covariance matrix also depends on the parameters, and needs to be re-computed
for every sampled model \citep{2017MNRAS.472.4244H}.

The proposed solution to this challenge is to use an Approximate Bayesian Computation (ABC) framework.
To generate model predictions from a multivariate distribution is possible
under a singular covariance, and does not require the precision matrix. We
consider three examples with increasing complexity:
an affine function with diagonal data covariance, a non-linear model with
correlated data points inspired by the cosmological weak-gravitational-lensing
power spectrum, and a realistic weak-lensing case. The results demonstrate that the proposed ABC approach can recover the input parameter values in these
inference problems, with as low a number of simulations as $n_{\textrm{s}} = 2$.

Other methods that reduce the number 
of required simulations to $n_{\textrm{s}} < p$ add an apriori known covariance component,
for example
shrinkage methods \citep{2017MNRAS.466L..83J,2019MNRAS.483..189H} or other hybrid approaches \citep{2018MNRAS.473.4150F}.
These methods were
tested for $p/n_{\rm s} \leq 12$, while the proposed ABC method can provide reliable parameter constraints
for a covariance matrices up to $p /
n_{\textrm{s}} = 375$.

Data compression has been suggested as a way forward to reduce the requirements
on simulations,
e.g.~\citet{2000MNRAS.317..965H,2014arXiv1409.0863A,2018MNRAS.476L..60A,2018PhRvD..97h3004C,2021MNRAS.501..954J}.
Some of these methods however require the computation of the precision matrix of the
un-compressed data. In addition, data compression can potentially lose important information. The proposed ABC
approach can work with compressed data, but does not depend on it.

In our simulation study, we found that the estimated parameter means, their standard error, and the standard deviations of their variance
do not depend on $n_{\textrm s}$. This is in contrast to MCMC sampling, where the 
standard errors of the estimates display a bias that depends on the number of simulations.
The standard errors of the parameter estimates from ABC are in most cases larger than the
Cram\'er-Rao lower bound.
Using the predictions from a Hotelling $T^2$ distribution, corresponding to an estimated covariance matrix that follows a Wishart distribution, we obtain similar standard errors with ABC.

We introduce a new distance function based on the autocorrelation
function (acf) of the data (Eq.~\ref{eq:d_acf}), which does not use the covariance matrix. The proposed acf distance can account for correlations in the data without relying on a precision matrix.
Overall, the proposed ABC framework provides a strategy for inference of correlated observations when it is not possible to run numerous $N$-body simulations to estimate the precision matrix.

\section*{Acknowledgements}

The authors thank the anonymous referee for a thorough and careful review, which helped us to improve the manuscript.
We would like to thank Christian Robert, Elena Sellentin, and Andy Taylor for useful discussions. EEOI was financially supported by CNRS as part of its MOMENTUM programme over the 2018 -- 2020 period.
We gratefully acknowledge support from the CNRS/IN2P3 Computing Center (Lyon - France) for providing computing and data-processing resources needed for this work.
This research made use of the software packages
Astropy\footnote{\url{http://www.astropy.org}}, a community-developed core Python package for Astronomy \citep{astropy:2013, astropy:2018},
 Pystan \citep{pystan}, Scipy \citep{scipy}, and Statsmodel \citep{seabold2010statsmodels}.

%%%%%%%%%%%%%%%%%%%%%%%%%%%%%%%%%%%%%%%%%%%%%%%%%%

\begin{appendix}

\section{Parameter covariance matrix for the multivariate normal distribution}
\label{sec:phi}

\subsection{Estimated parameter covariance and its covariance matrix}
\label{sec:sd_var_fisher}

The $n_\theta \times p$ matrix $\mat M$ defined in Eq.~\eqref{eq:fish}
has rank $n_\theta$. The parameter covariance matrix is the inverse of the Fisher matrix,
$\mat \Phi = \mat F^{-1}.$
Since in our case $n_\theta = 2$, we can analytically take the inverse of $\mat F$ to obtain the parameter variance
\begin{equation}
    \textrm{Var}\left( \hat \theta_i \right)
        = \Phi_{ii} = \left| \mat F \right|^{-1}
        F_{jj},
    \label{eq:phi_ii}
\end{equation}
for $(i, j) \in \{(1, 2), (2, 1)\}$.

An estimator of $\mat \Phi$ can be defined using the estimated precision matrix,
$\hat{\mat \Phi} = \left[ \mat M^\textrm{t} \hat \Psi \mat M \right]^{-1} = \left[ \mat M^\textrm{t} (\hat {\mat C})^{-1} \mat M \right]^{-1}$. Since $\nu \hat{\mat C} \sim {\cal W}_p(C, \nu)$, we can apply Theorem 3.3.13 from \cite{gupta1999matrix}, according to which the estimated parameter covariance matrix follows a Wishart distribution with scale matrix $\mat \Phi = \left[ \mat M^\textrm{t} {\mat C}^{-1} \mat M \right]^{-1}$ and degrees of freedom $\nu - p + n_\theta$:
$\left( \nu - p + n_\theta \right) \hat{\mat \Phi} \sim {\cal W}_{n_\theta}(\mat \Phi, \nu - p + n_\theta)$. Following Table \ref{tab:estimators}, we can write the covariance of the parameter covariance as
\begin{equation}
    \mathbb{E} \left( \hat \Phi_{ij} - \Phi_{ij} \right)^2
        = \left( \nu - p + n_\theta \right)^{-1}
        \left( \Phi_{ij}^2 - \Phi_{ii} \Phi_{jj} \right)^2
    \label{eq:SDVarF}
\end{equation}
Note that different expression have been obtained in \cite{2014MNRAS.442.2728T} and \cite{2017MNRAS.464.4658S}.

The following two sections compute the numerical values of the parameter variances for the affine function (Sect.~\ref{sec:toy_example}), and the weak-lensing inspired example (Sect.~\ref{sec:WL_example}), respectively.

\subsection{Standard errors for the affine function}
\label{sec:Fisher_matrix_ex_1}

The following expressions correspond to the example of an affine function model $\vec m(\vec \theta) = a \vec x + b \vec 1$, described in Sect.~\ref{sec:toy_example}.
With $\theta_1 = a$ and $\theta_2 = b$, thus
$\partial \vec m / \partial \theta_1 = \vec x$ and
$\partial \vec m / \partial \theta_2 = \vec 1$,
the Fisher matrix is written as
\begin{align}
	\mat F = &
    	\left(
    	\begin{array}{cc}
        	\vec x^{\rm t} \mat \Psi \vec x &
            \vec x^{\rm t} \mat \Psi \vec 1 \\
            \vec x^{\rm t} \mat \Psi \vec 1 &
            \vec 1^{\rm t} \mat \Psi \vec 1
    	\end{array}
        \right)
        \,
        =
        \frac 1 {\sigma^{2}}
        \left(
    	\begin{array}{cc}
        \sum_i x_i^2 &
        \sum_i x_i \\
        \sum_i x_i &
        p
    	\end{array}
        \right)
        .
        \label{eq:F_affine}
\end{align}
Using the expected values of the uniformly-distributed $x_i$ and their squares, we obtain
$F_{11} = (p \Delta^2) / (12 \sigma^2)$, and $F_{12} = F_{21} = 0$.
The parameter estimate variances are then
\begin{equation}
	\textrm{Var}(\hat a) = \Phi_{11} = \left({\mat F}^{-1}\right)_{11} = \frac{\sigma^2}{p} \frac{12}{\Delta^2}; \quad
   \textrm{Var}(\hat b) = \Phi_{22} = ({\mat F}^{-1})_{22} = \frac{\sigma^2}{p}.
    \label{eq:Phi_ana}
\end{equation}
With $\Delta = 200, p = 750$, and $\sigma^2 = 5$,
the numerical values for the standard errors are SE$(\hat a)
= 1 / (500 \sqrt{2}) \approx 0.001414$, SE$(\hat b)
= 1/(5\sqrt{6}) \approx 0.08165$.

\subsection{Standard errors for the weak-lensing inspired example}
\label{sec:app_phi_WL_example}

The following expressions correspond to the example of the weak-lensing inspired example defined in Sect.~\ref{sec:WL_example}.
The model function $\vec m$ is given by
Eqs.~\eqref{eq:qu}, \eqref{eq:u0_c}, and \eqref{eq:y_tm3}. We first compute the derivatives analytically,
\begin{align}
  \frac{\partial \vec m}{\partial t} = & \frac{a \log 10}{2 \times 6.05} (\vec x - x_0 \vec 1) \odot \vec m(\vec \theta) ; \nonumber  \\
  \frac{\partial \vec m}{\partial A} = & \frac{2}{A} \vec m(\vec \theta),
\end{align}
where $\odot$ denotes Hadamard (element-wide) multiplication.
The parameter variances are then obtained by numerically computing the Fisher matrix Eq.~\eqref{eq:fish}, and taking the inverse Eq.~\eqref{eq:phi_ii}.

\subsection{Standard errors for the realistic weak-lensing inspired}

For the realistic weak-lensing example (Sect.~\ref{sec:WL_model}), the
derivatives of the data vector $y\Sim$ with respect to the parameters
$\Omega_\textrm{m}$ and $\sigma_8$ are computed numerically using finite differences. The parameter variances are then obtained following
\ref{sec:app_phi_WL_example}.

\end{appendix}

%%%%%%%%%%%%%%%%%%%% REFERENCES %%%%%%%%%%%%%%%%%%

%\bibliographystyle{imsart-nameyear}
\bibliographystyle{elsarticle-harv}
\bibliography{ref}

\end{document}